\documentclass[aip,amsmath,amssymb,rsi,groupedaddress]{revtex4-1}
\usepackage{graphicx}
\usepackage{epsfig}
\usepackage{amsmath}
\usepackage{graphics}
\usepackage{epsfig}
\usepackage{yfonts}
\usepackage{amsmath}
\usepackage{amsfonts}
\usepackage{amssymb}
\usepackage{xcolor}
\allowdisplaybreaks

\begin{document}

\title{Non-equilibrium Green's function theory for non-adiabatic effects in quantum transport: inclusion of electron-electron interactions}

\author{Vincent F. Kershaw and Daniel S. Kosov } 

\address{College of Science and Engineering, James Cook University, Townsville, QLD, 4811, Australia }

\begin{abstract}
Non-equilibrium Green's function theory for non-adiabatic effects in quantum transport  [Kershaw and Kosov, J.Chem. Phys. 2017, 147, 224109 and J. Chem. Phys.  2018, 149, 044121] is extended to the case of interacting electrons. We consider a general problem of quantum transport of interacting electrons through a central region with dynamically changing geometry. The approach is based on the separation of time scales in the non-equilibrium Green's functions and the use of the Wigner transformation to solve the Kadanoff-Baym equations.  The Green's functions and correlation self-energy are non-adiabatically expanded up to the second order central time derivatives. We produce expressions for Green's functions with non-adiabatic corrections and a  modified formula for electric current; both depend not only on instantaneous molecular junction geometry but also on nuclear velocities and accelerations.  The theory is illustrated by the study of electron transport through a model single-resonant level  molecular junction  with local electron-electron repulsion and a dynamically changing geometry.

\end{abstract}

\maketitle

\section{INTRODUCTION}
The description of correlated quantum many-body systems far away from equilibrium remains one of the most challenging problems of physics.\cite{negf} A molecular junction, a single molecule attached to two macroscopic leads held at different chemical potentials, represents an ultimate challenge for theory due to two distinct and interconnected  features: electron-electron interactions and structural flexibility.\cite{moletronics} Molecular junctions carry  an extremely large current density of the order of microamperes per square nm, many orders of magnitude larger than is usual in mesoscopic devices. Molecular junctions also show a  high inhomogeneity of electron density such as cusps on nuclear positions, lone pairs and electron concentrations along chemical bonds. Compared to traditional semiconductor devices, the correct treatment of electron-electron interactions is critical for determining electrical conductivity.\cite{PhysRevB.75.075102,PhysRevB.77.115333,PhysRevB.79.155110,PhysRevB.83.115108,thygesen07,sandalov03,fransson05,galperin-hubbard08,galperin-hubbard09,dzhioev11a,dzhioev11b,dzhioev12,rabani13,cohen2015,PhysRevB.88.235426,PhysRevLett.101.066804,wang13,PhysRevB.79.245125,PhysRevB.80.165305} The  electronic correlations are amplified in far-from-equilibrium situations owing to  an increased number of available states for electron-electron scattering. 
\\
Unlike silicon based devices, the molecular junctions are not rigid structures  as they are always prone to large amplitude nuclear motions, current induced conformational changes and even chemical reactions.\cite{seideman:521,Pistolesi08,fuse,dzhioev11,catalysis12,darwish2016,PhysRevLett.78.4410,doi:10.1021/jacs.6b10700,C2CP40851A,PhysRevLett.107.046801,doi:10.1021/acs.jpclett.8b00940,dundas12,thoss11,peskin18} These characteristic properties of molecular junctions have proven difficult to implement theoretically. Different models have been developed to tackle this problem in a wide range of theoretical formalisms: master equations, \cite{PhysRevB.69.245302,fcblockade05,PhysRevB.83.115414,may02,PhysRevB.94.201407,segal15,kosov17-wtd,kosov17-nonren,dzhioev14,dzhioev15,kosov18} path integral,\cite{rabani08} scattering theory, \cite{ness05,PhysRevB.70.125406,peskin07,doi:10.1063/1.3231604}  non-equilibrium Green's functions,\cite{ryndyk06,dahnovsky:014104,galperin06,ryndyk07,hartle08,rabani14,hartle15,PhysRevB.75.205413} and multilayer multiconfiguration time-dependent Hartree theory.\cite{wang11,wang13,wang13b,Wang2016117} These approaches  (apart from several recent works \cite{Pistolesi08,fuse,dzhioev11,bode12,catalysis12,galperin15,subotnik17-prl,kershaw17,kershaw18}) have almost universally assumed that nuclear vibrations are harmonic in nature, effectively describing systems that deviate slightly from their zero-current and equilibrium geometry. In addition, it is common to treat either the electron-vibration or molecule-electrode couplings as the small parameter  relative to other energy scales  in the system.

Moreover, in standard approaches, the electronic correlations and dynamical conformational changes are often neglected altogether (as is done in most computer codes for first principles electron transport calculations). \cite{reimers2000,Evers04,Brandbyge02,Taylor01,li:1347,thygesen:111,Calzolari2004,Fujimoto03,Xue02,smeagol} At best they are considered as separate and uncoupled entities: either electronic-correlations are treated for a "frozen" molecular geometry\cite{PhysRevB.75.075102,PhysRevB.77.115333,PhysRevB.79.155110,PhysRevB.83.115108,thygesen07,sandalov03,fransson05,galperin-hubbard08,galperin-hubbard09,dzhioev11a,dzhioev11b,dzhioev12,rabani13,cohen2015,PhysRevB.88.235426,PhysRevLett.101.066804,wang13,PhysRevB.79.245125,PhysRevB.80.165305} or molecular conformation changes are considered for non-interacting resonant-level model.\cite{dundas12,fuse,catalysis12} The scope of current theoretical work with simultaneous modeling electronic correlations and nuclear dynamics is, at present, limited and all these considering nuclear motion as harmonic vibrations around equilibrium geometry.\cite{paaske05,koch05,galperin:035301}

Recently, several authors have used gradient expansions,\cite{rammer-book} a technique reaching back to work of Kadanoff and Baym,\cite{kadanoff-baym} in their theoretical approaches, \cite{bode12,fuse,catalysis12,PhysRevB.92.235440,galperin15,subotnik17-prl,kershaw17,subotnik18,kershaw18} where classically described nuclei are treated as a slow disturbance in non-equilibrium Green's functions. Our previous works expanded on this methodology, where we developed a transport theory in the non-equilibrium Green's function formalism that takes into account the non-adiabatic effects of nuclear motion. \cite{kershaw17, kershaw18} The theory allowed for the computation of non-adiabatic effects associated with nuclear motion in the central region and at the molecule-electrode interface. This paper serves as a natural extension, where we further develop the theory to account for electron-electron interactions in the system. Equations are resolved by treating nuclear velocity as the small parameter through a gradient expansion where, consequently, one can separate an adiabatic component originating from a frozen nuclear geometry and a non-adiabatic term that arises in nuclear motions. As such, the equations can be solved while making no assumptions regarding small or harmonic nuclear motion, nor is it required that electron-nuclear, molecule-electrode or electron-electron interactions be considered small.

The outline of the paper is as follows. Section II describes the general theory:  solution of the real-time Kadanoff-Baym equations for molecular advanced, retarded and lesser Green's functions using Wigner representation. This section also contains the derivation of the general expression for electron current which includes electronic correlations along with the non-adiabatic corrections due to nuclear dynamical motion. In section III we illustrate the proposed theory by the application to electron transport through a molecular junction modelled by the Anderson model with a time-dependent energy level. We use atomic units in the derivations throughout the paper ($\hbar= |e|= m_e= 1$).
 
\section{THEORY}
\subsection{Hamiltonian, Green's Functions and Self-Energies}
In this section we present the governing Hamiltonian, Green's functions and self-energies, and introduce notation conventions that will be used throughout the document. We start with the generic Hamiltonian for quantum transport through the system which consists of a central (scattering) region connected to two macroscopic leads. We will call the central region a "molecule", but all our results are applicable to the general case where  the central region are represented by a quantum dot, atom, or  any other nano-scale system with time varying geometry.
This Hamiltonian takes the form
\begin{equation}
H=H_M + H_L +H_R +H_{ML} + H_{ML},
\label{hamiltonian}
\end{equation}
where $H_M$ is the Hamiltonian for the molecule, $H_L$  and  $H_R$ are the Hamiltonians for the left and  right leads,  $H_{ML}$  and $H_{MR}$ are  the interaction between the central region and the left and right leads, respectively. The Hamiltonian for the central region  is explicitly time-dependent and contains electronic interaction
\begin{equation}
H_M= \sum_{ij} h_{ij} (\mathbf R(t)) a^\dag_i a_j + V_C,
\end{equation}
where we have made use of the multidimensional vector $\mathbf R(t)$ that describes the geometry (positions of nuclei in the case of molecular junction) of the central region at time $t$. The quantity $h_{ij} (\mathbf R(t))$ is matrix element of the single-particle part of the molecular Hamiltonian  computed in some basis. The electron-electron interaction or, possibly, the electron-phonon interaction if we choose to treat part of the nuclear degrees of freedom not included in $\mathbf R(t)$ quantum mechanically is described by $V_C$.  As is typical for molecules, we assume that only the single-particle part of the Hamiltonian depends explicitly on molecular geometry.
We omit, at this point, the classical part of the Hamiltonian which generates trajectory $ \mathbf R(t)$ since it does not influence equations for the electronic Green's functions. 

The left and right leads  are modelled as macroscopic reservoirs of non-interacting electrons
\begin{equation}
H_L +H_R =  \sum_{ k\nu=L,R}  \varepsilon_{k\alpha}  a^{\dagger}_{k\alpha} a_{k\alpha},
\end{equation}
where $a^{\dagger}_{k\nu}$($a_{k\nu}$) creates (annihilates) an electron in the single-particle state $k$ of either the left ($\nu=L$) or  the right ($\nu=R$) electrode.
  The coupling between central region and left   and right leads  are described by the tunnelling interaction
\begin{equation}
H_{ML}+H_{MR}=  \sum_{i k \nu=L,R } (t_{k\nu i} a^{\dagger}_{k \nu} a_i +\mbox{h.c.} ),
\end{equation}
where $t_{k\nu i}$ is the tunnelling amplitudes between  leads and molecular single-particle states. Next, we define the exact (non-adiabatic, computed with full time-dependent Hamiltonian along a given nuclear trajectory $\mathbf R(t)$) retarded, advanced, and lesser Green's functions as
\begin{equation}
{\cal G}_{ij}^R(t,t') = -i \theta(t-t') \langle \{a_i (t), a^\dag_j (t')\} \rangle,
\end{equation}
\begin{equation}
{\cal G}_{ij}^A(t,t') = [ {\cal G}_{ij}^R(t',t) ]^*
\end{equation}
and
\begin{equation}
{\cal G}_{ij}^<(t,t') = i \langle a^\dag_j (t')a_i (t) \rangle.
\end{equation}
The self-energies of leads are not affected by the time-dependent molecular Hamiltonian and they are defined in the standard way.\cite{haug-jauho} Self-energy components for the leads are given by
\begin{equation}
{\Sigma}_{\nu ij}^R(t,t') = -i \theta(t-t') \sum_{k} t^*_{k\nu  i}  e^{-i \epsilon_{k \nu} (t-t')} t_{k \nu j},
\end{equation}
\begin{equation}
{\Sigma}_{\nu ij}^A(t,t') = [ {\Sigma}_{\nu ij}^R(t',t) ]^*
\end{equation}
and
\begin{equation}
{\Sigma}_{\nu ij}^<(t,t') =  i  \sum_{k} f_{\nu}(\epsilon_{k\nu}) t^*_{k\nu i}  e^{-i \epsilon_{k \nu} (t-t')} t_{k \nu j}.
\end{equation}
Here $f_{\nu}$ is the Fermi-Dirac distribution of the $\nu$  lead. The total lead self-energies are the sum of contributions from the left and right leads respectively as
\begin{equation}
{\Sigma}_{ij}^{R,A,<}(t,t') = {\Sigma}_{L ij}^{R,A,<}(t,t') + {\Sigma}_{R ij}^{R,A,<}(t,t').
\end{equation}
The retarded self-energies in the energy domain  are defined in a standard way as a Fourier transformation of time domain self-energies defined above:
\begin{equation}
\Sigma_{\nu ij}^R(\omega)= \Delta_{\nu ij }(\omega)  -\frac{i}{2} \Gamma_{\nu ij}(\omega),
\end{equation}
where the level-width functions are
\begin{equation}
 \Gamma_{\nu ij } (\omega) = 2 \pi \sum_{k} \delta(\omega-\epsilon_{k\nu})  t^*_{k\nu i} t_{k\nu j},
\end{equation}
and level-shift functions $\Delta_{\nu ij }(\omega)$ can be computed from $\Gamma_{\nu ij}(\omega)$  via Kramers-Kronig relation.\cite{haug-jauho} The advanced and lesser self-energies are computed from the retarded self-energy as
\begin{equation}
\Sigma_{\nu ij}^A(\omega) = (\Sigma_{\nu ji}^R(\omega))^*
\end{equation}
 and 
\begin{equation}
 \Sigma_{\nu ij}^<(\omega)= f_{\nu}(\omega)\left( \Sigma_{\nu ij}^A(\omega) - \Sigma_{\nu ij}^R(\omega)\right)= i f_{\nu}(\omega) \Gamma_{\nu ij}(\omega).
\end{equation}

\subsection{Separation of time-scales in the Kadanoff-Baym equations of motion}
We begin with the Kadanoff-Baym equations of motion for the retarded, advanced and lesser Green's functions \cite{haug-jauho}
\begin{multline}
\Big( i \partial_t - h(t) \Big){\cal G}^{<}(t,t') = \int dt_1 \Big[ \Big( \Sigma^R_C(t,t_1) + \Sigma^{R}(t,t_{1}) \Big) {\cal G}^<(t_1,t') \\
+ \Big( \Sigma^<_C(t,t_1) + \Sigma^{<}(t,t_{1}) \Big) {\cal G}^A(t_1,t') \Big]
\label{eom2}
\end{multline}
and 
\begin{equation}
\Big( i  \partial_t  - h(t) \Big) {\cal G}^{R/A}(t,t') = 
 \delta(t-t') +  \int dt_1 \Big( \Sigma^R_C(t,t_1) + \Sigma^{R/A}(t,t_{1}) \Big) {\cal G }^{R/A}(t_{1},t'),
 \label{eom1}
\end{equation}
where we consider the retarded and advanced equations collectively. Note that we have chosen to work with the Kadanoff-Baym equations in matrix form where the Green's functions, self-energies and Hamiltonian $h$ become matrices in the molecular orbital space. Here $\Sigma_C$ is the self-energy from correlation in the central region (also matrix in molecular orbital space) and the choice of the particular form  is not relevant for our immediate discussion.

We are motivated to transform the equations of motion to the Wigner space where fast and slow time scales are easily identifiable. To do so, we define the central time $T$ and relative time $\tau$ parameters 
\begin{equation}
T = \frac{1}{2}(t+t')
\end{equation}
and
\begin{equation}
\tau = t - t^{\prime},
\end{equation}
and introduce the Wigner representation of the Green's function
\begin{equation}
{\cal \widetilde G}(\omega,T) = \int  d\tau \; e^{i \omega \tau} {\cal G} (t,t') .
\end{equation}
The inverse transformation from the Wigner representation to time domain is
\begin{equation}
{\cal G} (t,t')= \frac{1}{2 \pi} \int  d\omega \; e^{-i \omega \tau}  {\cal \widetilde G}(\omega,T).
\end{equation}
It can be shown that applying the the Wigner transform to both sides of equations (\ref{eom1}) and (\ref{eom2}) yields the equations of motion in the Wigner space 
\begin{equation}
\Big( \omega  +\frac{i}{2}\partial_T  - e^{\frac{1}{2i} \partial_{T}^{h} \partial_{\omega}^{{\cal G }}}h (T) \Big) \widetilde{{\cal G }}^{R}(T,\omega) \\
= I +  e^{\frac{1}{2i}( \partial^{\Sigma}_T \partial^{\cal G}_\omega- \partial^{\Sigma}_\omega \partial^{\cal G}_T) } \Big( \widetilde{\Sigma}^R_C(T,\omega)+ \widetilde{\Sigma}^{R}(\omega) \Big) \widetilde{{\cal G }}^{R}(T,\omega)
\label{eomW1}
\end{equation}
and
\begin{multline}
\Big( \omega  +\frac{i}{2}\partial_T  - e^{\frac{1}{2i} \partial_{T}^{h} \partial_{\omega}^{{\cal G }}} h (T) \Big) \widetilde{{\cal G }}^{<}(T,\omega) = e^{\frac{1}{2i}( \partial^{\Sigma}_T \partial^{\cal G}_\omega- \partial^{\Sigma}_\omega \partial^{\cal G}_T) } \Big( \widetilde{\Sigma}^R_C(T,\omega)+ \widetilde{\Sigma}^{R}(\omega) \Big) \widetilde{{\cal G }}^{<}(T,\omega)
\\
+
e^{\frac{1}{2i}( \partial^{\Sigma}_T \partial^{\cal G}_\omega- \partial^{\Sigma}_\omega \partial^{\cal G}_T) } \Big( \widetilde{\Sigma}^<_C(T,\omega)+ \widetilde{\Sigma}^{<}(\omega) \Big) \widetilde{{\cal G }}^{A}(T,\omega).
\label{eomW2}
\end{multline}
Here $\partial^\Sigma$, $\partial^{\cal G }$ and $\partial^{h }$ mean the derivatives acting on the self-energy $\Sigma^R$, Green's function $\widetilde{{\cal G }}^{R}$ and Hamiltonian matrix $h$, respectively, with $I$ being the identity matrix. In order to make expressions more manageable we now drop the function notation and define the 'total' self-energy term $\widetilde{\Sigma}_{tot} = \widetilde{\Sigma} + \widetilde{\Sigma}_{C}$ to produce
\begin{equation}
\Big( \omega  +\frac{i}{2}\partial_T  - e^{\frac{1}{2i} \partial_{T}^{h} \partial_{\omega}^{{\cal G }}}h \Big) \widetilde{{\cal G }}^{R/A} \\
= I +  e^{\frac{1}{2i} ( \partial^{\Sigma}_T \partial^{\cal G}_\omega - \partial^{\Sigma}_\omega \partial^{\cal G}_T )} \widetilde{\Sigma}^{R/A}_{tot} \widetilde{{\cal G }}^{R/A}
\label{yep}
\end{equation}
and
\begin{equation}
\Big( \omega  +\frac{i}{2}\partial_T  - e^{\frac{1}{2i} \partial_{T}^{h} \partial_{\omega}^{{\cal G }}} h \Big) \widetilde{{\cal G }}^{<} 
= e^{\frac{1}{2i} ( \partial^{\Sigma}_T \partial^{\cal G}_\omega - \partial^{\Sigma}_\omega \partial^{\cal G}_T ) } \Big( \widetilde{\Sigma}^R_{tot} \widetilde{{\cal G }}^{<} +  \widetilde{\Sigma}^{<}_{tot} \widetilde{{\cal G }}^{A} \Big).
\label{no}
\end{equation}
So far no approximations have been made in regard to equations (\ref{yep}) and (\ref{no}). They describe the exact non-adiabatic evolution of the retarded, advanced and lesser Green's functions and must be solved along a given trajectory $\mathbf R(t)$ of nuclear coordinates. 

Working in the Wigner space allows us to naturally identify variation in $T$ as the small parameter of our theory by separating slow nuclear time-scales from the fast electronic time-scales. The slow nuclear motion is reflected by slow variation with central time $T$ and fast oscillations with relative time $\tau$ of the Green's functions.\cite{kershaw17}

Before expanding the exponential operators $e^{\frac{1}{2i} \partial_{T}^{h} \partial_{\omega}^{{\cal G }}}$  and $e^{\frac{1}{2i}( \partial^{\Sigma}_T \partial^{\cal G}_\omega- \partial^{\Sigma}_\omega \partial^{\cal G}_T) }$ in the equations of motion, we first survey the form of the central time derivatives of the Hamiltonian matrix and specify notation. 
The nuclear coordinates are represented as a $3N$-dimensional vector
\begin{equation}
\mathbf R=(\mathbf R_1, \mathbf R_2, ... , \mathbf R_N) = (x_1, x_2, x_3,  x_4, x_5, x_6, .....,  x_{3N-2}, x_{3N-1}, x_{3N}).
\end{equation}
The first and second order time derivatives of the single-particle  Hamiltonian matrix are
\begin{equation}
\partial_T h(\mathbf R) =   \dot{ x}_\alpha \Lambda_\alpha (\mathbf R)
\end{equation}
and
\begin{equation}
\partial^2_T h(\mathbf R)
=  \ddot{x}_\alpha  \Lambda_\alpha (\mathbf R)+  \dot{x}_\alpha \dot{x}_{\beta}  \Phi_{\alpha \beta} (\mathbf R),
\end{equation}
where we assumed a summation over the repeated Greek indices and 
introduced matrices in the molecular orbital space 
\begin{equation}
\Lambda_{\alpha}(\mathbf R)=\frac{\partial h(\mathbf R)}{\partial x_\alpha}
\end{equation}
and
\begin{equation}
\Phi_{\alpha \beta}(\mathbf R) =\frac{\partial^2 h(\mathbf R)}{\partial x_\alpha \partial x_\beta}.
\end{equation}
The presence of the exponential operators in the Wigner space form of the Kadanoff-Baym equations makes finding a solution a difficult problem. A natural next step would be to expand the exponential operators $e^{\frac{1}{2i} \partial_{T}^{h} \partial_{\omega}^{{\cal G }}}$ and $e^{\frac{1}{2i} ( \partial^{\Sigma}_T \partial^{\cal G}_\omega - \partial^{\Sigma}_\omega \partial^{\cal G}_T )}$ as a MacLaurin series. Acting derivative terms in the resultant MacLaurin series on the Green's functions results in powers of $\frac{\Omega}{\Gamma}$, where $\Omega$ is the characteristic frequency of a molecular vibration and $\Gamma$ is the molecular level broadening. For example, if we consider $e^{\frac{1}{2i} \partial_{T}^{h} \partial_{\omega}^{{\cal G }}}$ acting on $\widetilde{\mathcal{G}}^{R}$, the first term in the exponential series acting on the retarded Green's function will be of the order of $ \dot{x} \partial_{\omega} \widetilde{{\cal G}}^{R} \sim (\frac{\Omega}{\Gamma})$ and the second term will be of order $ \ddot{x} \partial_{\omega}^{2} \widetilde{{\cal G}}^{R} \sim (\frac{\Omega}{\Gamma})^{2}$. By limiting our model to consider situations where the characteristic timescale of nuclear vibrations is large relative to the electron tunnelling time, effectively meaning that $\frac{\Omega}{\Gamma} <1$, we expand the MacLaurin series of $e^{\frac{1}{2i} \partial_{T}^{h} \partial_{\omega}^{{\cal G }}}$ and $e^{\frac{1}{2i} ( \partial^{\Sigma}_T \partial^{\cal G}_\omega - \partial^{\Sigma}_\omega \partial^{\cal G}_T )}$ and keep the first three terms in this expansion.

Implementing this assumption, the resultant equations of motion in the Wigner space take the form
\begin{multline}
\Big( \omega  +\frac{i}{2}\partial_T - h - \frac{1}{2i} \dot{x}_{\alpha} \Lambda_{\alpha} \partial_{\omega} + \frac{1}{8} \ddot{x}_\alpha \Lambda_\alpha  \partial^{2}_{\omega} + \frac{1}{8} \dot{x}_{\alpha} \dot{x}_{\beta} \Phi_{\alpha \beta}  \partial^{2}_{\omega} \Big) \widetilde{{\cal G }}^{R/A} \\ = I + \Big( I + \frac{1}{2i} \partial^{\Sigma}_{T} \partial^{{\cal G}}_{\omega} - \frac{1}{2i} \partial^{\Sigma}_{\omega} \partial^{{\cal G}}_{T} - \frac{1}{8} (\partial^{\Sigma}_{T} \partial^{{\cal G}}_{\omega} )^2+ \frac{1}{4} \partial^{\Sigma}_{T \omega} \partial^{{\cal G}}_{T \omega} - \frac{1}{8} (\partial^{\Sigma}_{\omega} \partial^{{\cal G}}_{T})^2 \Big) \widetilde{\Sigma}^{R/A}_{tot} \widetilde{{\cal G }}^{R/A}
\end{multline}
and
\begin{multline}
\Big( \omega  +\frac{i}{2}\partial_T - h - \frac{1}{2i} \dot{x}_{\alpha} \Lambda_{\alpha} \partial_{\omega} + \frac{1}{8} \ddot{x}_\alpha \Lambda_\alpha  \partial^{2}_{\omega} +  \frac{1}{8} \dot{x}_{\alpha} \dot{x}_{\beta} \Phi_{\alpha \beta}  \partial^{2}_{\omega} \Big) \widetilde{{\cal G }}^{<} \\ =  \Big( I + \frac{1}{2i} \partial^{\Sigma}_{T} \partial^{{\cal G}}_{\omega} - \frac{1}{2i} \partial^{\Sigma}_{\omega} \partial^{{\cal G}}_{T} - \frac{1}{8} (\partial^{\Sigma}_{T} \partial^{{\cal G}}_{\omega} )^2+ \frac{1}{4} \partial^{\Sigma}_{T \omega} \partial^{{\cal G}}_{T \omega} - \frac{1}{8} (\partial^{\Sigma}_{\omega} \partial^{{\cal G}}_{T})^2 \Big) \Big( \widetilde{\Sigma}^R_{tot} \widetilde{{\cal G }}^{<} +  \widetilde{\Sigma}^{<}_{tot} \widetilde{{\cal G }}^{A} \Big).
\end{multline}
The following frequently appearing quantities are symbolically redefined as
\begin{equation}
\mathcal{A}^{R/A} = I - \partial_{\omega}\widetilde{\Sigma}^{R/A}_{tot},
\label{A_bar}
\end{equation}
\begin{equation}
\mathcal{B}^{R/A} = \dot{x}_{\alpha} \Lambda_{\alpha} + \partial_{T}\widetilde{\Sigma}^{R/A}_{tot}
\label{B_bar}
\end{equation}
and
\begin{equation}
\mathcal{C}^{R/A} = \ddot{x}_{\alpha} \Lambda_{\alpha} + \dot{x}_{\alpha} \dot{x}_{\beta} \Phi_{\alpha \beta} + \partial_{T}^{2}\widetilde{\Sigma}^{R/A}_{tot}.
\label{C_bar}
\end{equation}
Note that the quantity $\mathcal{B}^{R/A}$ is summed over the repeated Greek index $\alpha$ and the quantity $\mathcal{C}^{R/A}$ is summed over the indices $\alpha$ and $\beta$. Rearranging the equations slightly and implementing the above quantities we find
\begin{multline}
\Big( \omega - h - \widetilde{\Sigma}^{R/A}_{tot} \Big) \widetilde{{\cal G }}^{R/A} = I + \frac{1}{2i} \Big( \mathcal{B}^{R/A} \partial_{\omega} + \mathcal{A}^{R/A} \partial_{T} \Big) \widetilde{{\cal G }}^{R/A} \\ - \frac{1}{8} \Big( \mathcal{C}^{R/A} \partial^{2}_{\omega} - 2 \partial_{T \omega} \widetilde{\Sigma}^{R/A}_{tot} \partial_{T \omega} +  \partial^{2}_{\omega} \widetilde{\Sigma}^{R/A}_{tot} \partial^{2}_{T} \Big) \widetilde{{\cal G }}^{R/A}
\end{multline}
and
\begin{multline}
\Big( \omega - h - \widetilde{\Sigma}^R_{tot} \Big) \widetilde{{\cal G }}^{<} = \widetilde{\Sigma}^{<}_{tot} \widetilde{{\cal G }}^{A} + \frac{1}{2i} \Big( \mathcal{B}^{R} \partial_{\omega} + \mathcal{A}^{R} \partial_{T} \Big) \widetilde{{\cal G }}^{<} + \frac{1}{2i} \Big( \partial_{T} \widetilde{\Sigma}^{<}_{tot} \partial_{\omega} - \partial_{\omega} \widetilde{\Sigma}^{<}_{tot} \partial_{T} \Big) \widetilde{{\cal G }}^{A} \\ - \frac{1}{8} \Big( \mathcal{C}^{R} \partial^{2}_{\omega} - 2 \partial_{T \omega} \widetilde{\Sigma}^{R}_{tot} \partial_{T \omega} +  \partial^{2}_{\omega} \widetilde{\Sigma}^{R}_{tot} \partial^{2}_{T} \Big) \widetilde{{\cal G }}^{<} - \frac{1}{8} \Big( \partial^{2}_{T} \widetilde{\Sigma}^{<}_{tot} \partial^{2}_{\omega} - 2 \partial_{T \omega} \widetilde{\Sigma}^{<}_{tot} \partial_{T \omega} +  \partial^{2}_{\omega} \widetilde{\Sigma}^{<}_{tot} \partial^{2}_{T} \Big) \widetilde{{\cal G }}^{A}.
\end{multline}
We observe that the RHS of the equations of motion contain first and second order derivatives acted on from the left by matrix quantities (self-energy components and their derivatives). It is useful to define the differential operators $\hat{P}_{i}^{g}$ and $\hat{D}_{i}^{g}$ which consist of first and second order derivatives respectively. We choose to denote the self-energy components of a given differential operator with the superscript $g \in \{ R,A,< \}$. The operators $\hat{P}_{i}^{g}$ and $\hat{D}_{i}^{g}$ are defined as 
\begin{equation}
\hat{P}_{i}^{g} = \frac{1}{2i} \Big( X_{i}^{g} \partial_{\omega} + Y_{i}^{g} \partial_{T} \Big)
\end{equation}
and
\begin{equation}
\hat{D}_{i}^{g} = - \frac{1}{8} \Big( X_{i}^{g} \partial^{2}_{\omega} - 2 Y_{i}^{g} \partial_{T \omega} +  Z_{i}^{g} \partial^{2}_{T} \Big). 
\end{equation}
We note that we have made use of the subscript $i \in \{ a , b \}$ which differentiates between the two first and second order operators with different matrix coefficients. For example, when considering the first order differential operator, the subscript $i$ differentiates between
\begin{equation}
\hat{P}^{g}_{a} = \frac{1}{2i} \Big( \mathcal{B}^{g} \partial_{\omega} + \mathcal{A}^{g} \partial_{T} \Big)
\end{equation}
and 
\begin{equation}
\hat{P}^{g}_{b} = \frac{1}{2i} \Big( \partial_{T} \widetilde{\Sigma}^{g}_{tot} \partial_{\omega} - \partial_{\omega} \widetilde{\Sigma}^{g}_{tot} \partial_{T} \Big).
\end{equation}
Similarly, when considering the second order differential operator the subscript $i$ differentiates between 
\begin{equation}
\hat{D}^{g}_{a} =  - \frac{1}{8} \Big( \mathcal{C}^{g} \partial^{2}_{\omega} - 2 \partial_{T \omega} \widetilde{\Sigma}^{g}_{tot} \partial_{T \omega} +  \partial^{2}_{\omega} \widetilde{\Sigma}^{g}_{tot} \partial^{2}_{T} \Big)
\end{equation}
and
\begin{equation}
\hat{D}^{g}_{b} = - \frac{1}{8} \Big( \partial^{2}_{T} \widetilde{\Sigma}^{g}_{tot} \partial^{2}_{\omega} - 2 \partial_{T \omega} \widetilde{\Sigma}^{g}_{tot} \partial_{T \omega} + \partial^{2}_{\omega} \widetilde{\Sigma}^{g}_{tot} \partial^{2}_{T} \Big). 
\end{equation}
These definitions are useful as it allows us to take derivatives of a Green's function component once for arbitrary matrix quantities with the resultant expression being adapted depending on what matrix quantities are being considered (see Appendix). Using these simplification schemes we find that the equations of motion become
\begin{equation}
\Big( \omega - h - \widetilde{\Sigma}^{R/A}_{tot} \Big) \widetilde{{\cal G }}^{R/A} = I + \Big( \hat{P}_{a}^{R/A} + \hat{D}_{a}^{R/A} \Big) \widetilde{{\cal G }}^{R/A}
\end{equation}
and
\begin{equation}
\Big( \omega - h - \widetilde{\Sigma}^R_{tot} \Big) \widetilde{{\cal G }}^{<} = \widetilde{\Sigma}^{<}_{tot} \widetilde{{\cal G }}^{A} + \Big( \hat{P}_{a}^{R} + \hat{D}_{a}^{R} \Big) \widetilde{{\cal G }}^{<} + \Big( \hat{P}_{b}^{<} + \hat{D}_{b}^{<} \Big) \widetilde{{\cal G }}^{A}.
\end{equation}
The equations of motion have now been transformed into the Wigner space and a more manageable form. There are two intricacies associated with a perturbative solution to the equations of motion: dealing with (i) the functional dependencies of $\widetilde{\Sigma}_{tot}$ on the Green's function components in the system space and (ii) the Green's function components explicit in the equations of motion. We will deal with these intricacies separately where we first deal with the functional dependencies of the self-energy terms as seen in the next section.

\subsection{Non-adiabatic expansion of the correlation self-energy}
The self-energy terms have explicit dependencies on the Green's functions. We start by taking a second order perturbation in the Green's functions 
\begin{equation}
\widetilde{\mathcal{G}} = \widetilde{G}_{(0)} + \lambda \widetilde{G}_{(1)} + \lambda^2 \widetilde{G}_{(2)}.
\end{equation}
Here $\lambda$ is the "book-keeping" parameter to keep track of  the orders of central time derivatives in the derivations. Terms that are linear in $\lambda$ will be also linear in central-time derivatives and  terms containing $\lambda^2$ will be quadratic in central-time derivatives in the equations that follow. The parameter $\lambda$ will be set to $1$ in the end of the derivations. It follows that the correlation self-energy components are approximated as 
\begin{equation}
\widetilde{\Sigma}_C \big[ \widetilde{\mathcal{G}} \big] = \widetilde{\Sigma}_C \big[ \widetilde{G}_{(0)} + \lambda \widetilde{G}_{(1)} + \lambda^2 \widetilde{G}_{(2)} \big],
\end{equation}
which we reform as 
\begin{equation}
\widetilde{\Sigma}_C \big[ \widetilde{\mathcal{G}} \big] = \widetilde{\Sigma}_C \big[ \widetilde{G}_{(0)} + \lambda \big( \widetilde{G}_{(1)} + \lambda \widetilde{G}_{(2)} \big) \big] = \widetilde{\Sigma}_C \big[ \widetilde{G}_{(0)} + \lambda \widetilde{D} \big],
\end{equation}
by introducing the quantity $\widetilde{D} = \widetilde{G}_{(1)} + \lambda \widetilde{G}_{(2)}$ which encompasses non-adiabatic corrections. We now take a functional Taylor series expansion in the self-energy to get
\begin{multline}
\widetilde{\Sigma}_C \big[ \widetilde{\mathcal{G}} \big] = \underbrace{\widetilde{\Sigma}_C \big[ \widetilde{G}_{(0)} \big]}_{\widetilde{\Sigma}_{T(0)}} + \underbrace{\int dT^{\prime} d\omega^{\prime} \frac{\delta \widetilde{\Sigma}_C \big[ \widetilde{G}_{(0)} \big]}{\delta \widetilde{G}_{(0)}} \delta \widetilde{D}(T^{\prime},\omega^{\prime})}_{\widetilde{\Sigma}_{C(1)}} \\ + \underbrace{\int dT^{\prime} d\omega^{\prime} dT^{\prime \prime} d\omega^{\prime \prime} \frac{\delta^{2} \widetilde{\Sigma}_C \big[ \widetilde{G}_{(0)} \big]}{\delta \widetilde{G}_{(0)} \delta \widetilde{G}_{(0)}} \delta \widetilde{D}(T^{\prime},\omega^{\prime}) \delta \widetilde{D}(T^{\prime \prime},\omega^{\prime \prime})}_{\widetilde{\Sigma}_{C(2)}},
\end{multline}
where we have labeled the adiabatic and non-adiabatic components respectively. We have made use of the quantity $\delta \widetilde{D}$ where it follows by definition of $\widetilde{D}$ that $\delta \widetilde{D} = \delta \widetilde{G}_{(1)} + \lambda \delta \widetilde{G}_{(2)}$. It follows that the total self-energy is then given by
\begin{equation}
\widetilde{\Sigma}_{tot} \big[ \widetilde{\mathcal{G}} \big] = \widetilde{\Sigma} + \sum_{j=0}^{2} \widetilde{\Sigma}_{Cj},
\end{equation}
where $j$ keeps track of perturbative corrections where $j \in \{ (0),(1),(2) \}$. The specification of the perturbed self-energy components allows for the representation of the remaining matrix quantities $\mathcal{A}^{R/A}$, $\mathcal{B}^{R/A}$ and $\mathcal{C}^{R/A}$ present in the equations of motion as
\begin{equation}
\mathcal{A}^{R/A} = \underbrace{I - \partial_{\omega} \widetilde{\Sigma}^{R/A}_{tot(0)}}_{\mathcal{A}^{R/A}_{(0)}} - \underbrace{\partial_{\omega} \widetilde{\Sigma}^{R/A}_{tot(1)}}_{\mathcal{A}^{R/A}_{(1)}} - \underbrace{\partial_{\omega} \widetilde{\Sigma}^{R/A}_{tot(2)}}_{\mathcal{A}^{R/A}_{(2)}},
\end{equation}
\begin{equation}
\mathcal{B}^{R/A} = \underbrace{\dot{x}_{\alpha} \Lambda_{\alpha} + \partial_{T} \widetilde{\Sigma}^{R/A}_{tot(0)}}_{\mathcal{B}^{R/A}_{(1)}} + \underbrace{\partial_{T} \widetilde{\Sigma}^{R/A}_{tot(1)}}_{\mathcal{B}^{R/A}_{(2)}}
\end{equation}
and 
\begin{equation}
\mathcal{C}^{R/A} = \underbrace{\ddot{x}_{\alpha} \Lambda_{\alpha} + \dot{x}_{\alpha} \dot{x}_{\beta} \Phi_{\alpha \beta} + \partial^{2}_{T} \widetilde{\Sigma}^{R/A}_{tot(0)}}_{\mathcal{C}^{R/A}_{(2)}},
\end{equation} 
where above we have labeled the adiabatic and non-adiabatic components. Note that we have neglected terms above the second order after applying the central time derivatives as they do not appear in the solution. As such $\mathcal{A}^{R/A}$, $\mathcal{B}^{R/A}$ and $\mathcal{C}^{R/A}$ can be represented as the sum of their adiabatic and non-adiabatic components as 
\begin{equation}
\mathcal{A}^{R/A} = \sum^{2}_{j=0} \mathcal{A}^{R/A}_{j},
\end{equation}
\begin{equation}
\mathcal{B}^{R/A} = \sum^{2}_{j=1} \mathcal{B}^{R/A}_{j}
\end{equation}
and 
\begin{equation}
\mathcal{C}^{R/A} = \mathcal{C}^{R/A}_{(2)}.
\end{equation} 
It is useful to define the perturbative orders of the differential operators $\hat{P}$ and $\hat{D}$ using the expressions derived above. Considering the first order differential operators and substituting in explicit expressions we find that
\begin{equation}
\hat{P}_{a}^{g} = \underbrace{\frac{1}{2i} \Big( \mathcal{B}^{g}_{(1)} \partial_{\omega} + \mathcal{A}^{g}_{(0)} \partial_{T} \Big)}_{\hat{P}^{g}_{a(1)}} + \underbrace{ \frac{1}{2i} \Big( \mathcal{B}^{g}_{(2)} \partial_{\omega} + \mathcal{A}^{g}_{(1)} \partial_{T} \Big)}_{\hat{P}^{g}_{a(2)}}
\end{equation}
and 
\begin{equation}
\hat{P}_{b}^{g} = \underbrace{\frac{1}{2i} \Big( \partial_{T} \widetilde{\Sigma}^{g}_{tot(0)} \partial_{\omega} - \partial_{\omega} \widetilde{\Sigma}^{g}_{tot(0)} \partial_{T} \Big)}_{\hat{P}^{g}_{b(1)}} + \underbrace{ \frac{1}{2i} \Big( \partial_{T} \widetilde{\Sigma}^{g}_{tot(1)} \partial_{\omega} - \partial_{\omega} \widetilde{\Sigma}^{g}_{tot(1)} \partial_{T} \Big)}_{\hat{P}^{g}_{b(2)}}.
\end{equation}
Considering now the second order differential operators we find that 
\begin{equation}
\hat{D}^{g}_{a} = \underbrace{ - \frac{1}{8} \Big( \mathcal{C}^{g}_{(2)} \partial^{2}_{\omega} - 2 \partial_{T \omega} \widetilde{\Sigma}^{g}_{tot(0)} \partial_{T \omega} +  \partial^{2}_{\omega} \widetilde{\Sigma}^{g}_{tot(0)} \partial^{2}_{T} \Big) }_{\hat{D}^{g}_{a(2)}}
\end{equation}
and
\begin{equation}
\hat{D}^{g}_{b} = \underbrace{ - \frac{1}{8} \Big( \partial^{2}_{T} \widetilde{\Sigma}^{g}_{tot(0)} \partial^{2}_{\omega} - 2 \partial_{T \omega} \widetilde{\Sigma}^{g}_{tot(0)} \partial_{T \omega} +  \partial^{2}_{\omega} \widetilde{\Sigma}^{g}_{tot(0)} \partial^{2}_{T} \Big) }_{\hat{D}^{g}_{b(2)}}.
\end{equation}
In all equations above we have labeled the perturbative orders and neglected terms that exceed the second order. This convention is very useful as it allows us to resolve the equations of motion into its perturbative orders easily, along with simplifying notation. As a result we find that the equations of motion, with some rearrangement, become 
\begin{equation}
\Big( \omega - h - \widetilde{\Sigma}^{R/A}_{tot(0)} \Big) \widetilde{\mathcal{G}}^{R/A} = I + \Big( \widetilde{\Sigma}^{R/A}_{tot(1)} + \widetilde{\Sigma}^{R/A}_{tot(2)} \Big) \widetilde{\mathcal{G}}^{R/A} + \Big( \hat{P}^{R/A}_{a(1)} + \hat{P}^{R/A}_{a(2)} + \hat{D}^{R/A}_{a(2)} \Big) \widetilde{\mathcal{G}}^{R/A}
\end{equation}
and
\begin{multline}
\Big( \omega - h - \widetilde{\Sigma}^{R}_{tot(0)} \Big) \widetilde{\mathcal{G}}^{<} = \Big( \widetilde{\Sigma}^{R}_{tot(1)} + \widetilde{\Sigma}^{R}_{tot(2)} \Big) \widetilde{\mathcal{G}}^{<} + \Big( \widetilde{\Sigma}^{R}_{tot(0)} + \widetilde{\Sigma}^{R}_{tot(1)} + \widetilde{\Sigma}^{R}_{tot(2)} \Big) \widetilde{\mathcal{G}}^{A} \\ + \Big( \hat{P}^{R}_{a(1)} + \hat{P}^{R}_{a(2)} + \hat{D}^{R}_{a(2)} \Big) \widetilde{\mathcal{G}}^{<} + \Big( \hat{P}^{<}_{b(1)} + \hat{P}^{<}_{b(2)} + \hat{D}^{<}_{b(2)} \Big) \widetilde{\mathcal{G}}^{A}.
\end{multline}
The equations of motion have been altered as a consequence of dealing with the self-energy dependencies on Green's functions in the system space. In the next section we deal with the Green's function explicit in the equations of motion themselves.

\subsection{Solution of the Kadanoff-Baym equation}
In the previous section, the Wigner space Kadanoff-Baym equations have been derived retaining terms in the central time derivatives up to the second order. We will now solve these equations. We will first consider the retarded and advanced Green's function components before considering the lesser  Green's function. Derivations from this point will make frequent use of commutator and anti-commutator operations given by $[\cdot, \cdot]_{-}$ and $[\cdot, \cdot]_{+}$ respectively. 
\subsubsection{Retarded/Advanced Components}
\label{RAsection}
We now take a second order perturbation in the retarded/advanced Keldysh components such that
\begin{equation}
\widetilde{\mathcal{G}}^{R/A} = \widetilde{G}^{R/A}_{(0)} + \lambda \widetilde{G}^{R/A}_{(1)} + \lambda^2 \widetilde{G}^{R/A}_{(2)}
\end{equation}
where $\lambda$ is the "book-keeping" parameter for the orders of central time derivatives. We substitute the perturbative expansion of $\widetilde{\mathcal{G}}^{R/A}$ into the retarded/advanced equations of motion to find
\begin{multline}
\Big( \omega - h - \widetilde{\Sigma}^{R/A}_{tot(0)} \Big) \Big( \widetilde{G}^{R/A}_{(0)} + \lambda \widetilde{G}^{R/A}_{(1)} + \lambda^2 \widetilde{G}^{R/A}_{(2)} \Big) = I + \Big( \widetilde{\Sigma}^{R/A}_{tot(1)} + \widetilde{\Sigma}^{R/A}_{tot(2)} \Big) \Big( \widetilde{G}^{R/A}_{(0)} + \lambda \widetilde{G}^{R/A}_{(1)} \\ + \lambda^2 \widetilde{G}^{R/A}_{(2)} \Big) + \Big( \hat{P}^{R/A}_{a(1)} + \hat{P}^{R/A}_{a(2)} + \hat{D}^{R/A}_{a(2)}  \Big) \Big( \widetilde{G}^{R/A}_{(0)} + \lambda \widetilde{G}^{R/A}_{(1)} + \lambda^2 \widetilde{G}^{R/A}_{(2)} \Big).
\end{multline}
We now split the retarded/advanced equations of motion based on order to get (letting $\lambda = 0$) 
\begin{equation}
\Big( \omega - h - \widetilde{\Sigma}^{R/A}_{tot(0)} \Big) \widetilde{G}^{R/A}_{(0)} = I,
\label{zerothR/A}
\end{equation}
\begin{equation}
\Big( \omega - h - \widetilde{\Sigma}^{R/A}_{tot(0)} \Big) \widetilde{G}^{R/A}_{(1)} = \widetilde{\Sigma}^{R/A}_{tot(1)} \widetilde{G}^{R/A}_{(0)} + \hat{P}^{R/A}_{a(1)} \widetilde{G}^{R/A}_{(0)} 
\label{firstR/A}
\end{equation}
and 
\begin{multline}
\Big( \omega - h - \widetilde{\Sigma}^{R/A}_{tot(0)} \Big) \widetilde{G}^{R/A}_{(2)}  = \widetilde{\Sigma}^{R/A}_{tot(1)} \widetilde{G}^{R/A}_{(1)} + \widetilde{\Sigma}^{R/A}_{tot(2)} \widetilde{G}^{R/A}_{(0)} + \hat{P}^{R/A}_{a(1)} \widetilde{G}^{R/A}_{(1)} + \hat{P}^{R/A}_{a(2)} \widetilde{G}^{R/A}_{(0)} \\ + \hat{D}^{R/A}_{a(2)} \widetilde{G}^{R/A}_{(0)}.
\label{secondR/A}
\end{multline}
It is easy to see that (\ref{zerothR/A}) can be solved to give 
\begin{equation}
\widetilde{G}^{R/A}_{(0)} = \Big( \omega - h - \widetilde{\Sigma}^{R/A}_{tot(0)} \Big)^{-1},
\end{equation}
which is the standard and well-understood adiabatic retarded/advanced Green's function which we re-label to 
\begin{equation}
G^{R/A} = \widetilde{G}^{R/A}_{(0)}.
\end{equation}
We now consider (\ref{firstR/A}) where we first rearrange in terms of $\widetilde{G}^{R/A}_{(1)}$ to get 
\begin{equation}
\widetilde{G}^{R/A}_{(1)} = G^{R/A} \widetilde{\Sigma}^{R/A}_{tot(1)} G^{R/A} + G^{R/A} \hat{P}^{R/A}_{a(1)} G^{R/A}.
\end{equation} 
It can easily be shown that 
\begin{equation}
\partial_{\omega} G^{R/A} = - G^{R/A} \Big( I - \partial_{\omega} \widetilde{\Sigma}^{R/A}_{tot(0)} \Big) G^{R/A}  = - G^{R/A} \mathcal{A}_{(0)}^{R/A} G^{R/A}
\label{div1}
\end{equation}
and 
\begin{equation}
\partial_{T} G^{R/A} = G^{R/A} \Big( \dot{x}_{\alpha} \Lambda_{\alpha} + \partial_{T} \widetilde{\Sigma}^{R/A}_{tot(0)} \Big) G^{R/A} = G^{R/A} \mathcal{B}_{(1)}^{R/A} G^{R/A}.
\label{div2}
\end{equation}
As a result we can show that 
\begin{multline}
\hat{P}^{R/A}_{a} G^{R/A} = \frac{1}{2i} \Big( \mathcal{A}_{(0)}^{R/A} G^{R/A} \mathcal{B}^{R/A}_{(1)} G^{R/A} - \mathcal{B}^{R/A}_{(1)} G^{R/A}\mathcal{A}^{R/A}_{(0)} G^{R/A} \Big) \\ = \frac{1}{2i} \Big[ \mathcal{A}_{(0)}^{R/A} G^{R/A}, \mathcal{B}^{R/A}_{(1)} G^{R/A} \Big]_{-},
\end{multline}
where this allows us to specify $\widetilde{G}^{R/A}_{(1)}$ as 
\begin{equation}
\widetilde{G}^{R/A}_{(1)} = G^{R/A} \widetilde{\Sigma}^{R/A}_{tot(1)} G^{R/A} + \frac{1}{2i} G^{R/A} \Big[ \mathcal{A}_{(0)}^{R/A} G^{R/A}, \mathcal{B}^{R/A}_{(1)} G^{R/A} \Big]_{-}.
\label{retarded1}
\end{equation} 
Now focusing on the equation of motion for $\widetilde{G}^{R/A}_{(2)}$ given by (\ref{secondR/A}), we rearrange in terms of $\widetilde{G}^{R/A}_{(2)}$ to get 
 \begin{multline}
\widetilde{G}^{R/A}_{(2)}  = G^{R/A} \widetilde{\Sigma}^{R/A}_{tot(1)} \widetilde{G}^{R/A}_{(1)} + G^{R/A} \widetilde{\Sigma}^{R/A}_{tot(2)} G^{R/A} + G^{R/A} \hat{P}^{R/A}_{a(2)} G^{R/A} + G^{R/A} \hat{D}^{R/A}_{a(2)} G^{R/A} \\ + G^{R/A} \hat{P}^{R/A}_{a(1)} \widetilde{G}^{R/A}_{(1)}.
\label{retarded2}
\end{multline}
Using (\ref{div1}), (\ref{div2}) and the definition of $\hat{P}^{R/A}_{a(2)}$, one can show that
\begin{equation}
\hat{P}^{R/A}_{a(2)} G^{R/A} = \frac{1}{2i} \Big( \mathcal{A}_{(1)}^{R/A} G^{R/A} \mathcal{B}^{R/A}_{(1)} G^{R/A} - \mathcal{B}^{R/A}_{(2)} G^{R/A}\mathcal{A}^{R/A}_{(0)} G^{R/A} \Big).
\end{equation}
We now consider the second order derivatives of $G^{R/A}$ in order to compute an expression for $\hat{D}^{R/A}_{a(2)} G^{R/A}$. These derivatives are given by 
\begin{equation}
\partial^{2}_{\omega} G^{R/A} = 2 G^{R/A} \Big( \mathcal{A}^{R/A}_{(0)} G^{R/A} \Big)^2 + G^{R/A} \partial^{2}_{\omega} \widetilde{\Sigma}^{R/A}_{tot(0)} G^{R/A}, 
\end{equation}
\begin{equation}
\partial_{T \omega} G^{R/A} = - G^{R/A} \Big[ \mathcal{B}^{R/A}_{(1)} G^{R/A}, \mathcal{A}^{R/A}_{(0)} G^{R/A} \Big]_{+} + G^{R/A} \partial_{T \omega} \widetilde{\Sigma}^{R/A}_{tot(0)} G^{R/A}  
\end{equation}
and 
\begin{equation}
\partial^{2}_{T} G^{R/A} = 2 G^{R/A} \Big( \mathcal{B}^{R/A}_{(1)} G^{R/A} \Big)^2 + G^{R/A} \partial^{2}_{T} \widetilde{\Sigma}^{R/A}_{tot(0)} G^{R/A}.
\end{equation}
These derivatives allow us to show that 
\begin{multline}
\hat{D}^{R/A}_{a(2)} G^{R/A} =  - \frac{1}{4} \Big[ \mathcal{C}^{R/A} G^{R/A} \Big( \mathcal{A}^{R/A}_{(0)} G^{R/A} \Big)^2 + \partial_{T \omega} \widetilde{\Sigma}^{R/A}_{tot(0)} G^{R/A} \Big[ \mathcal{B}^{R/A}_{(1)} G^{R/A}, \mathcal{A}^{R/A}_{(0)} G^{R/A} \Big]_{+} \\ + \partial^2_{\omega} \widetilde{\Sigma}^{R/A}_{tot(0)} G^{R/A} \Big( \mathcal{B}^{R/A}_{(1)} G^{R/A} \Big)^2 \Big] - \frac{1}{8} \Big[ \mathcal{C}^{R/A} G^{R/A} \partial^{2}_{\omega} \widetilde{\Sigma}^{R/A}_{tot(0)} G^{R/A} - 2 \Big( \partial_{T \omega} \widetilde{\Sigma}^{R/A}_{tot(0)} G^{R/A} \Big)^2 \\ + \partial^2_{\omega} \widetilde{\Sigma}^{R/A}_{tot(0)} G^{R/A} \partial^{2}_{T} \widetilde{\Sigma}^{R/A}_{tot(0)} G^{R/A} \Big].
\end{multline}
One can compute explicit expressions for the derivative $\hat{P}^{R/A}_{a(1)} \widetilde{G}^{R/A}_{(1)}$ in the term $G^{R/A} \hat{P}^{R/A}_{a(1)} \widetilde{G}^{R/A}_{(1)}$ which, in the interest of presentation, has been left to the Appendix B. Ultimately, this leads to an explicit expression for $\widetilde{G}^{R/A}_{(2)}$, which has been relegated to Appendix A. 

\subsubsection{Lesser Components}
\label{lessersection}
We now turn our attention to deriving non-adiabatic  corrections to the lesser Green's function, which, as we shall find, is significantly more tedious. Taking a second order perturbation of the lesser Green's function component as 
\begin{equation}
\widetilde{\mathcal{G}}^{<} = \widetilde{G}^{<}_{(0)} + \lambda \widetilde{G}^{<}_{(1)} + \lambda^2 \widetilde{G}^{<}_{(2)},
\end{equation}
which when substituted into the equation of motion for the lesser Green's function becomes
\begin{multline}
\Big( \omega - h - \widetilde{\Sigma}^R_{tot(0)} - \widetilde{\Sigma}^{R}_{tot(1)} - \widetilde{\Sigma}^{R}_{tot(2)} \Big) \Big( \widetilde{G}^{<}_{(0)} + \widetilde{G}^{<}_{(1)} + \widetilde{G}^{<}_{(2)} \Big) = \Big( \widetilde{\Sigma}^{<}_{tot(0)} + \widetilde{\Sigma}^{<}_{tot(1)} + \widetilde{\Sigma}^{<}_{tot(2)} \Big) \\ \times \Big( G^{A} + \widetilde{G}^{A}_{(1)} + \widetilde{G}^{A}_{(2)} \Big) +  \Big( \hat{P}^{R}_{a(1)} + \hat{P}^{R}_{a(2)} + \hat{D}^{R}_{a(2)} \Big) \Big( \widetilde{G}^{<}_{(0)} + \widetilde{G}^{<}_{(1)} + \widetilde{G}^{<}_{(2)} \Big) \\ + \Big( \hat{P}^{<}_{b(1)} + \hat{P}^{<}_{b(2)} + \hat{D}^{<}_{b(2)} \Big) \Big( G^{A} + \widetilde{G}^{A}_{(1)} + \widetilde{G}^{A}_{(2)} \Big). 
\end{multline}
We now split the lesser equation of motion based on order to get 
\begin{equation}
\Big( \omega - h - \widetilde{\Sigma}^{R}_{tot(0)} \Big) \widetilde{G}^{<}_{(0)} = \widetilde{\Sigma}^{<}_{tot(0)} G^{A},
\label{zerothlesser}
\end{equation}
\begin{equation}
\Big( \omega - h - \widetilde{\Sigma}^{R}_{tot(0)} \Big) \widetilde{G}^{<}_{(1)} = \widetilde{\Sigma}^{R}_{tot(1)} \widetilde{G}^{<}_{(0)} + \widetilde{\Sigma}^{<}_{tot(0)} \widetilde{G}^{A}_{(1)} + \widetilde{\Sigma}^{<}_{tot(1)} G^{A} + \hat{P}^{R}_{a(1)} \widetilde{G}^{<}_{(0)} + \hat{P}^{<}_{b(1)} G^{A}
\label{firstlesser}
\end{equation}
and 
\begin{multline}
\Big( \omega - h - \widetilde{\Sigma}^{R}_{tot(0)} \Big) \widetilde{G}^{<}_{(2)}  = \widetilde{\Sigma}^{R}_{tot(2)} \widetilde{G}^{<}_{(0)} + \widetilde{\Sigma}^{R}_{tot(1)} \widetilde{G}^{<}_{(1)} + \widetilde{\Sigma}^{<}_{tot(2)} G^{A} + \widetilde{\Sigma}^{<}_{tot(1)} \widetilde{G}^{A}_{(1)} + \widetilde{\Sigma}^{<}_{tot(0)} \widetilde{G}^{A}_{(2)} \\ + \hat{P}^{R}_{a(1)} \widetilde{G}^{<}_{(1)} + \Big( \hat{P}^{R}_{a(2)} + \hat{D}^{R}_{a(2)} \Big) \widetilde{G}^{<}_{(0)} + \hat{P}^{<}_{b(1)} \widetilde{G}^{A}_{(1)} + \Big( \hat{P}^{<}_{b(2)} + \hat{D}^{<}_{b(2)} \Big) G^{A}.
\end{multline}
It is easy to see that (\ref{zerothlesser}) is solved to give 
\begin{equation}
 \widetilde{G}^{<}_{(0)} = G^{R} \widetilde{\Sigma}^{<}_{tot(0)} G^{A},
\end{equation}
which is the standard adiabatic lesser Green's function which we relabel to 
\begin{equation}
G^{<} = \widetilde{G}^{<}_{(0)}.
\end{equation}
Considering now (\ref{firstlesser}), we first rearrange in terms of $\widetilde{G}^{<}_{(1)}$ to get 
\begin{equation}
\widetilde{G}^{<}_{(1)} = G^{R} \widetilde{\Sigma}^{R}_{tot(1)} G^{<} + G^{R} \widetilde{\Sigma}^{<}_{tot(0)} \widetilde{G}^{A}_{(1)} + G^{R} \widetilde{\Sigma}^{<}_{tot(1)} G^{A} + G^{R} \hat{P}^{R}_{a(1)} G^{<} + G^{R} \hat{P}^{<}_{b(1)} G^{A}.
\end{equation}
From the adiabatic lesser Green's function it follows that we can calculate its derivatives. We find that 
\begin{equation}
\partial_{\omega} G^{<} = - G^{R} \mathcal{A}^{R}_{(0)} G^{<} - G^{<}  \mathcal{A}^{A}_{(0)} G^{A} + G^{R} \partial_{\omega} \widetilde{\Sigma}^{<}_{tot(0)} G^{A}
\end{equation}
and
\begin{equation}
\partial_{T} G^{<} = G^{R} \mathcal{B}^{R}_{(1)} G^{<} + G^{<} \mathcal{B}^{A}_{(1)} G^{A} \\ + G^{R} \partial_{T} \widetilde{\Sigma}^{<}_{tot(0)} G^{A}.
\end{equation} 
This allows us to show that 
\begin{multline}
\hat{P}^{R}_{a(1)} G^{<} = \frac{1}{2i} \Big( \mathcal{A}_{(0)}^{R} G^{R} \mathcal{B}^{R}_{(1)} G^{<} + \mathcal{A}_{(0)}^{R} G^{<} \mathcal{B}^{A}_{(1)} G^{A} - \mathcal{B}_{(1)}^{R} G^{R} \mathcal{A}^{R}_{(0)} G^{<} - \mathcal{B}_{(1)}^{R} G^{<}  \mathcal{A}^{A}_{(0)} G^{A} \Big) \\ + \frac{1}{2i} \Big( \mathcal{B}_{(1)}^{R} G^{R} \partial_{\omega} \widetilde{\Sigma}^{<}_{tot(0)} G^{A} +  \mathcal{A}_{(0)}^{R} G^{R} \partial_{T} \widetilde{\Sigma}^{<}_{tot(0)} G^{A} \Big).
\end{multline}
From the previous section we can find that 
\begin{equation}
\hat{P}^{<}_{b(1)} G^{A}_{(0)} = -   \frac{1}{2i} \Big( \partial_{\omega} \widetilde{\Sigma}^{<}_{tot(0)} G^{A} \mathcal{B}^{A}_{(1)} G^{A} + \partial_{T} \widetilde{\Sigma}^{<}_{tot(0)} G^{A} \mathcal{A}^{A}_{(0)} G^{A} \Big).
\end{equation}
Taking note of these quantities and applying some rearrangement, we find that $\widetilde{G}^{<}_{(1)}$ is given by
\begin{multline}
\widetilde{G}^{<}_{(1)} = G^{R} \widetilde{\Sigma}^{R}_{tot(1)} G^{<} + G^{R} \widetilde{\Sigma}^{<}_{tot(0)} G^{A}_{(1)} + G^{R} \widetilde{\Sigma}^{<}_{tot(1)} G^{A} + \frac{1}{2i} G^{R} \Big( \mathcal{A}_{(0)}^{R} G^{R} \mathcal{B}^{R}_{(1)} G^{<} \\ + \mathcal{A}_{(0)}^{R} G^{<} \mathcal{B}^{A}_{(1)} G^{A} - \mathcal{B}_{(1)}^{R} G^{R} \mathcal{A}^{R}_{(0)} G^{<} - \mathcal{B}_{(1)}^{R} G^{<} \mathcal{A}^{A}_{(0)} G^{A} \Big) + \frac{1}{2i} G^{R} \Big( \mathcal{B}_{(1)}^{R} G^{R} \partial_{\omega} \widetilde{\Sigma}^{<}_{tot(0)} G^{A} \\ + \mathcal{A}_{(0)}^{R} G^{R} \partial_{T} \widetilde{\Sigma}^{<}_{tot(0)} G^{A} \Big) - \frac{1}{2i} G^{R} \Big( \partial_{\omega} \widetilde{\Sigma}^{<}_{tot(0)} G^{A} \mathcal{B}^{A}_{(1)} G^{A} + \partial_{T} \widetilde{\Sigma}^{<}_{tot(0)} G^{A} \mathcal{A}^{A}_{(0)} G^{A} \Big),
\label{lesser1}
\end{multline}
where we have chosen to keep $\widetilde{G}^{A}_{(1)}$, given by (\ref{retarded1}), as an input in order to keep expressions more manageable. We now focus on solving the second order lesser equation of motion. It is found that $\widetilde{G}^{<}_{(2)}$ is given by 
\begin{multline}
\widetilde{G}^{<}_{(2)}  = G^{R} \widetilde{\Sigma}^{R}_{tot(2)} G^{<} + G^{R} \widetilde{\Sigma}^{R}_{tot(1)} \widetilde{G}^{<}_{(1)} + G^{R} \widetilde{\Sigma}^{<}_{tot(2)} G^{A} + G^{R} \widetilde{\Sigma}^{<}_{tot(1)} \widetilde{G}^{A}_{(1)} + G^{R} \widetilde{\Sigma}^{<}_{tot(0)} \widetilde{G}^{A}_{(2)} \\ + G^{R} \hat{P}^{R}_{a(1)} \widetilde{G}^{<}_{(1)} + G^{R} \Big( \hat{P}^{R}_{a(2)} + \hat{D}^{R}_{a(2)} \Big) G^{<} + G^{R} \hat{P}^{<}_{b(1)} \widetilde{G}^{A}_{(1)} + G^{R} \Big( \hat{P}^{<}_{b(2)} + \hat{D}^{<}_{b(2)} \Big) G^{A}.
\end{multline}
The explicit expression for $\widetilde{G}^{<}_{(2)}$ is particularly cumbersome where, in the interest of presentation, we relegate some of the more complicated terms to the Appendix (these will be clearly labeled). Expressions for differential operators acting on adiabatic advanced Green's functions along with the first order differential operator acting on the adiabatic lesser Green's function can be easily adjusted from previous sections and will not be repeated here. Furthermore, expressions for $\widetilde{G}^{A}_{(1)}$, $\widetilde{G}^{A}_{(2)}$ and $\widetilde{G}^{<}_{(1)}$ will be left as inputs. Implementing the more manageable terms we find that $\widetilde{G}^{<}_{(2)}$ take the form 
\begin{multline}
\widetilde{G}^{<}_{(2)}  = G^{R} \widetilde{\Sigma}^{R}_{tot(2)} G^{<} + G^{R} \widetilde{\Sigma}^{R}_{tot(1)} \widetilde{G}^{<}_{(1)} + G^{R} \widetilde{\Sigma}^{<}_{tot(2)} G^{A} + G^{R} \widetilde{\Sigma}^{<}_{tot(1)} \widetilde{G}^{A}_{(1)} + G^{R} \widetilde{\Sigma}^{<}_{tot(0)} \widetilde{G}^{A}_{(2)} \\ + \frac{1}{2i} G^{R} \Big( \mathcal{A}_{(1)}^{R} G^{R} \mathcal{B}^{R}_{(1)} G^{<} + \mathcal{A}_{(1)}^{R} G^{<} \mathcal{B}^{A}_{(1)} G^{A} - \mathcal{B}_{(2)}^{R} G^{R} \mathcal{A}^{R}_{(0)} G^{<} - \mathcal{B}_{(2)}^{R} G^{<} \mathcal{A}^{A}_{(0)} G^{A} \Big) \\ + \frac{1}{2i} G^{R} \Big( \mathcal{B}_{(2)}^{R} G^{R} \partial_{\omega} \widetilde{\Sigma}^{<}_{tot(0)} G^{A} + \mathcal{A}_{(1)}^{R} G^{R} \partial_{T} \widetilde{\Sigma}^{<}_{tot(0)} G^{A} \Big) - \frac{1}{2i} G^{R} \Big( \partial_{\omega} \widetilde{\Sigma}^{<}_{tot(1)} G^{A} \mathcal{B}^{A}_{(1)} G^{A} \\ + \partial_{T} \widetilde{\Sigma}^{<}_{tot(1)} \mathcal{B}^{A}_{(2)} G^{A}\mathcal{A}^{A}_{(0)} G^{A} \Big) - \frac{1}{4} G^{R} \Big[ \partial^2_{T} \widetilde{\Sigma}^{<}_{tot(0)} G^{A} \Big( \mathcal{A}^{A}_{(0)} G^{A} \Big)^2 \\ + \partial_{T \omega} \widetilde{\Sigma}^{<}_{tot(0)} G^{A} \Big[ \mathcal{B}^{A}_{(1)} G^{A}, \mathcal{A}^{A}_{(0)} G^{A} \Big]_{+} + \partial^2_{\omega} \widetilde{\Sigma}^{<}_{tot(0)} G^{A} \Big( \mathcal{B}^{A}_{(1)} G^{A} \Big)^2 \Big] \\ - \frac{1}{8} G^{R} \Big[ \partial^2_{T} \widetilde{\Sigma}^{<}_{tot(0)} G^{A} \partial^{2}_{\omega} \widetilde{\Sigma}^{A}_{tot(0)} G^{A} - 2 \Big( \partial_{T \omega} \widetilde{\Sigma}^{<}_{tot(0)} G^{A} \Big)^2 + \partial^2_{\omega} \widetilde{\Sigma}^{A}_{tot(0)} G^{A} \partial^{2}_{T} \widetilde{\Sigma}^{A}_{tot(0)} G^{A} \Big] \\ + \underbrace{G^{R} \hat{D}^{R}_{a(2)} G^{<}}_{\text{Appendix \ref{Appendix A1}}} + \underbrace{G^{R} \hat{P}^{<}_{b(1)} \widetilde{G}^{A}_{(1)}}_{\text{Appendix \ref{Appendix A2}}} + \underbrace{G^{R} \hat{P}^{R}_{a(1)} \widetilde{G}^{<}_{(1)}}_{\text{Appendix \ref{Appendix A3}}},
\end{multline}
where we have labeled the final three terms according to where in the Appendix their respective expressions can be found.

\subsection{Electric current with non-adiabatic corrections}
Having obtained the non-adiabatic corrections to the retarded, advanced  and lesser Green's function we now derive the equation for the electric current using the Meir-Wingreen formula. We begin with the general expression for electric current flowing into the molecule from left/right leads at time $t$:\cite{haug-jauho}
\begin{eqnarray}
\mathcal{J}_{\nu}(t) =2 \int  dt_1
\text{ReTr} \Big\{ {\cal G}^<(t,t_1)\Sigma_{\nu}^A(t_1,t)
+ {\cal G}^R(t,t_1) \Sigma_{\nu}^<(t_1,t) \Big\},
\end{eqnarray}
where the trace is taken over the molecular orbital indices and the subscript $\nu \in \{ L, R \}$ denotes the left or right lead. Note that we have chosen the symbol $\mathcal{J}$ to denote the exact current which is in terms of the exact Green's functions. Transforming this equation to the Wigner space we find  
\begin{equation}
\widetilde{\mathcal{J}}_{\nu}(T,\omega) =2 e^{\frac{1}{2i}( \partial^{\cal G}_T \partial^{\Sigma}_\omega - \partial^{\cal G}_\omega \partial^{\Sigma}_T) } \text{Re} \text{Tr} \Big\{ \widetilde{\mathcal{G}}^<(T,\omega) \widetilde{\Sigma}_{\nu}^A(T,\omega) + \widetilde{\mathcal{G}}^R(T,\omega) \widetilde{\Sigma}_{\nu}^<(T,\omega) \Big\},
\end{equation}
where the real-time current can be computed as  
\begin{equation}
\mathcal{J}_{\nu}(t) =\frac{1}{\pi} \int d \omega e^{\frac{1}{2i}( \partial^{\cal G}_T \partial^{\Sigma}_\omega - \partial^{\cal G}_\omega \partial^{\Sigma}_T) } \text{Re} \text{Tr} \Big\{ \widetilde{\mathcal{G}}^<(T,\omega) \widetilde{\Sigma}_{\nu}^A(T,\omega) + \widetilde{\mathcal{G}}^R(T,\omega) \widetilde{\Sigma}_{\nu}^<(T,\omega) \Big\}.
\end{equation}
Expanding the exponential to the second order we get (dropping function dependencies)
\begin{multline}
\mathcal{J}_{\nu}(t) = \frac{1}{\pi} \int d \omega \text{ReTr} \Big\{ \widetilde{\mathcal{G}}^< \widetilde{\Sigma}_{\nu}^A + \widetilde{\mathcal{G}}^R \widetilde{\Sigma}_{\nu}^< + \frac{1}{2i} \Big( \partial_{T} \widetilde{\mathcal{G}}^< \partial_{\omega} \widetilde{\Sigma}_{\nu}^A + \partial_{T} \widetilde{\mathcal{G}}^R \partial_{\omega} \widetilde{\Sigma}_{\nu}^< \Big) \\ - \frac{1}{2i} \Big( \partial_{\omega} \widetilde{\mathcal{G}}^< \partial_{T} \widetilde{\Sigma}_{\nu}^A + \partial_{\omega} \widetilde{\mathcal{G}}^R \partial_{T} \widetilde{\Sigma}_{\nu}^< \Big) - \frac{1}{8} \Big( \partial_T^2 \widetilde{\mathcal{G}}^< \partial_\omega^2 \widetilde{\Sigma}_{\nu}^A  - 2 \partial_{T \omega} \widetilde{\mathcal{G}}^< \partial_{T \omega} \widetilde{\Sigma}_{\nu}^A + \partial_\omega^2 \widetilde{\mathcal{G}}^< \partial_T^2 \widetilde{\Sigma}_{\nu}^A \Big) \\ - \frac{1}{8} \Big( \partial_T^2 \widetilde{\mathcal{G}}^R \partial_\omega^2 \widetilde{\Sigma}_{\nu}^<  - 2 \partial_{T \omega} \widetilde{\mathcal{G}}^R \partial_{T \omega} \widetilde{\Sigma}_{\nu}^< + \partial_\omega^2 \widetilde{\mathcal{G}}^R \partial_T^2 \widetilde{\Sigma}_{\nu}^< \Big) \Big\}.
\label{perturb2}
\end{multline}
We now consider the previously computed perturbative expansions of the Green's functions in the system space which we implement by making the substitution
\begin{equation}
\widetilde{\mathcal{G}} = G + \lambda \widetilde{G}_{(1)} + \lambda^2 \widetilde{G}_{(2)},
\label{perturb}
\end{equation}
in the power of the smallness parameter. It follows that one can then compute the adiabatic lead current $J_{\nu}(t)$ and its non-adiabatic lead current corrections given by $J_{\nu(1)}(t)$ and $J_{\nu(2)}(t)$ such that 
\begin{equation}
\mathcal{J}_{\nu}(t) = J_{\nu}(t) + J_{\nu(1)}(t) + J_{\nu(2)}(t),
\label{currentall}
\end{equation}
where the symbol $J$ has been used to denote the non-exact current. Substituting (\ref{perturb}) into (\ref{perturb2}), setting the smallness parameter to $\lambda = 1$ and splitting the equation based on order, we find that the adiabatic lead current and its non-adiabatic lead current corrections are given by
\begin{equation}
J_{\nu}(t) = \frac{1}{\pi} \int d \omega \text{ReTr} \Big\{ G^< \widetilde{\Sigma}_{\nu}^A + G^R \widetilde{\Sigma}_{\nu}^< \Big\},
\label{current1}
\end{equation}
\begin{multline}
J_{\nu(1)}(t) = \frac{1}{\pi} \int d \omega \text{ReTr} \Big\{ \widetilde{G}_{(1)}^< \widetilde{\Sigma}_{\nu}^A + \widetilde{G}_{(1)}^R \widetilde{\Sigma}_{\nu}^< - \frac{1}{2i} \Big( \partial_{T \omega} G^< \widetilde{\Sigma}_{\nu}^A + \partial_{T \omega} G^R \widetilde{\Sigma}_{\nu}^< \Big) \\ - \frac{1}{2i} \Big( \partial_{\omega} G^< \partial_{T} \widetilde{\Sigma}_{\nu}^A + \partial_{\omega} G^R \partial_{T} \widetilde{\Sigma}_{\nu}^< \Big)\Big\}
\label{current2}
\end{multline}
and 
\begin{multline}
J_{\nu,(2)}(t) = \frac{1}{\pi} \int d \omega \text{ReTr} \Big\{ \widetilde{G}_{(2)}^< \widetilde{\Sigma}_{\nu}^A + \widetilde{G}_{(2)}^R \widetilde{\Sigma}_{\nu}^<  - \frac{1}{2i} \Big( \partial_{T \omega} \widetilde{G}_{(1)}^< \widetilde{\Sigma}_{\nu}^A + \partial_{T \omega} \widetilde{G}_{(1)}^R \widetilde{\Sigma}_{\nu}^< \Big) \\ - \frac{1}{2i} \Big( \partial_{\omega} \widetilde{G}_{(1)}^< \partial_{T} \widetilde{\Sigma}_{\nu}^A + \partial_{\omega} \widetilde{G}_{(1)}^R \partial_{T} \widetilde{\Sigma}_{\nu}^< \Big) - \frac{1}{8} \Big( \partial_{T \omega}^2 G^< \widetilde{\Sigma}_{\nu}^A + 2 \partial_{T} \partial^{2}_{\omega} G^< \partial_{T} \widetilde{\Sigma}_{\nu}^A + \partial_{\omega}^2 G^< \partial_T^2 \widetilde{\Sigma}_{\nu}^A \Big) \\ - \frac{1}{8} \Big( \partial_{T \omega}^2 G^R \widetilde{\Sigma}_{\nu}^<  + 2 \partial_{T} \partial^{2}_{\omega} G^R \partial_{T} \widetilde{\Sigma}_{\nu}^< + \partial_\omega^2 G^R \partial_T^2 \widetilde{\Sigma}_{\nu}^< \Big) \Big\}.
\label{current3}
\end{multline}
In the expressions above we have also made use of the identities 
\begin{equation}
\int d \omega \widetilde{A} \partial_{\omega} \widetilde{\Sigma} = - \int d \omega \partial_{\omega} \widetilde{A} \widetilde{\Sigma}
\end{equation}
and
\begin{equation}
\int d \omega \widetilde{A} \partial^2_{\omega} \widetilde{\Sigma} = \int d \omega \partial^2_{\omega} \widetilde{A} \widetilde{\Sigma}
\end{equation}
for arbitrary self-energy and Green's function quantities $\widetilde{\Sigma}$ and $\widetilde{A}$. It follows that equations (\ref{current1}), (\ref{current2}) and (\ref{current3}) combined according to (\ref{currentall}) allow one to compute the adiabatic current with non-adiabatic corrections provided that one has knowledge of the lead self-energy terms and the perturbative approximations to $\widetilde{\mathcal{G}}$.

\section{Model applications: Electron transport through Anderson impurity with time-dependent level energy}
\label{section3.1}
The proposed theory is exemplified by considering a time-dependent  model of a single electronic energy level with local electron-electron repulsion attached to two leads, the so called non-equilibrium Anderson impurity model. The time-dependent Anderson model has been used as an effective tool in the modeling of dynamical effects in molecular junctions; the problem  has been approached by a variety of methods such as non-equilibrium Monter-Carlo,\cite{Schmidt2008} time-dependend Hartree-Fock theory with vertex corrections,\cite{suzuki15} time-dependent non-crossing approximation\cite{Nordlander1999,Nordlander2000,Plihal2000} and time-dependent re-normalization group theory.\cite{Anders2005,Anders2006,Heidrich-Meisner2009}  The aforementioned works focused on either transient dynamics following sudden change of parameters or systems driven by periodic time dependent voltage bias.
The first case uses a time-independent Hamiltonian and the second employs Floquet theory to describe time dependence. In our work, the Hamiltonian has arbitrary time dependence.

The Hamiltonian for the molecule takes the form in second quantisation as
\begin{equation}
H_M= \epsilon(x) (a^\dag_{\downarrow} a_{\downarrow} + a^\dag_{\uparrow} a_{\uparrow} )+ V_C,
\end{equation}
where
\begin{equation}
V_C=U a^\dag_{\downarrow} a_{\downarrow} a^\dag_{\uparrow} a_{\uparrow}.
\end{equation}
The coordinate dependence of the resonant-level energy $\epsilon$ enters the problem in the following way. To compute the molecular non-adiabatic Green's functions at a given point $x=x_0$, we need to know the value of $\epsilon(x_0)$ as well as the first derivative $\Lambda =\epsilon'(x_0)$ and the second derivative $\Phi =\epsilon''(x_0)$.  In our calculations, we assume that $x_0$ correspond to the equilibrium molecular junction geometry. The resonant-level energy is taken to be aligned to the Fermi energy of the leads $\epsilon(x_0)=0$;   $\Lambda$ and $\Phi$  are considered as  parameters of the model which can be varied.

The molecular junction geometry undergoes stochastic thermal fluctuations around an equilibrium geometry. To account for this, we averaged the expressions for electric current using a Boltzmann velocity distribution with given temperature where, consequently, in the final expression for electric current we put $\langle \dot x \rangle = \langle \ddot x \rangle =0$ and $ \langle \dot{x}^2 \rangle \ne 0$. The range of physically relevant values for $\Lambda$, $\Phi$, and $ \langle \dot x \rangle$ are estimated in our previous paper\cite{kershaw17} based on quantum chemical calculations.

We now simplify the Green's functions and self-energy components by appealing to the spin degeneracy of the problem
(assuming a non-magnetic electronic population of the impurity). All cross-spin Green's functions vanish as
\begin{equation}
{\cal G}_{\uparrow  \downarrow} ( t, t') = {\cal G}_{\downarrow  \uparrow} (t,  t') = 0.
\label{nospin}
\end{equation}
The Green's functions for spin-up and spin-down electrons are identical
\begin{equation}
 {\cal G}_{\uparrow  \uparrow} ( t,   t')  ={\cal G}_{\downarrow  \downarrow}  ( t,  t') = {\cal G}(t,t'), 
\end{equation}
where in the last equality above we have introduced notation for Green's functions independent of the spin index, therefore the spin index can be omitted in most our derivations. The spin symmetry manifests itself in the lead self-energy components as
\begin{equation}
\Sigma_{\nu}^R= \Sigma_{\nu \uparrow \uparrow}^R=  \Sigma_{\nu \downarrow \downarrow}^R= - \frac{i}{2} \Gamma_{\nu},
\end{equation}
\begin{equation}
\Sigma_{\nu}^A= \Sigma_{\nu \uparrow \uparrow}^A=  \Sigma_{\nu \downarrow \downarrow}^A=  \frac{i}{2} \Gamma_{\nu}
\end{equation}
and
\begin{equation}
\Sigma_{\nu}^<(\omega)= \Sigma_{\alpha \uparrow \uparrow}^<(\omega)=  \Sigma_{\nu \downarrow \downarrow}^<(\omega) = i  f_{\nu}(\omega) \Gamma_{\nu}.
\end{equation}
Note that cross-spin self-energy components are  again zero and so will not be considered. 

We now consider  two-body interaction term $V_C$ and its corresponding correlation self-energy components where we choose to exemplify the proposed theory through the Hartree-Fock approximation. 
The correlation self-energy components then become 
\begin{equation}
\Sigma^{R/A}_{C \sigma \sigma}(t,t') = U n_{\sigma \sigma} (t) \delta(t-t') = - i U \mathcal{G}^{<}(t,t') \delta(t-t') 
\end{equation}
and 
\begin{equation}
\Sigma^<_{C \sigma \sigma}(t,t') = 0 ,
\end{equation}
where, once again, we see that the spin symmetry of the problem results in the correlation self-energies for the spin up and spin down processes being equal and the spin-transition components being zero. Observing that the retarded and advanced components for the correlation self-energies are equal, then we  simplify the notation to
\begin{equation}
\Sigma^R_{C \sigma \sigma}(t,t')=\Sigma^A_{C \sigma \sigma}(t,t') =\Sigma_C(t,t').
\end{equation}
Transforming the correlation self-energy to the Wigner space leads to 
\begin{equation}
\Sigma_C(T) =  - \frac{iU}{2 \pi}  \int d \omega \widetilde{\mathcal{G}}^<(T,\omega), 
\end{equation}
where we see that only functional dependencies on the central time are relevant. Taking a perturbative expansion in the Green's function above results in the series 
\begin{equation}
\Sigma_C(T) =  - \frac{iU}{2 \pi}  \int d \omega G^<(T,\omega) - \frac{i U}{2 \pi} \int d \omega \widetilde{G}_{(1)}^<(T,\omega) - \frac{i U}{2 \pi}  \int d \omega \widetilde{G}_{(2)}^<(T,\omega), 
\end{equation}
which becomes 
\begin{equation}
\Sigma_C(T) = - \underbrace{i U G^<(T)}_{\Sigma^{(0)}_{C}(T)} - \underbrace{i U  \widetilde{G}_{(1)}^<(T)}_{\Sigma^{(1)}_{C}(T)} - \underbrace{i U \widetilde{G}_{(2)}^<(T)}_{\Sigma^{(2)}_{C}(T)}.
\label{referee}
\end{equation}

Above perturbative orders have been identified which in turn allow us to determine the frequently appearing quantities ($\ref{A_bar}$), ($\ref{B_bar}$) and ($\ref{C_bar}$) simplify to the form (nuclear indices are absent in the single nuclear degree of freedom limit)
\begin{equation}
\mathcal{A}(T) = 1 = \mathcal{A}^{(0)}(T),
\end{equation}
\begin{equation}
\mathcal{B}(T) = \underbrace{\dot{x} \Lambda(T) - i U \partial_{T} G^<(T)}_{\mathcal{B}^{(1)}(T)} - \underbrace{i U \partial_{T} \widetilde{G}_{(1)}^<(T)}_{\mathcal{B}^{(2)}(T)}
\end{equation}
and 
\begin{equation}
\mathcal{C}(T) = \underbrace{\ddot{x} \Lambda(T) + \dot{x}^2 \Phi(T) - i U \partial^2_{T} G^<(T)}_{\mathcal{C}^{(2)}(T)}.
\end{equation}
Finalising the expressions above requires knowledge of the perturbative orders which will be the purpose of the next section. 

Following the perturbative expansion procedure detailed in the main body of the paper, it can be found that the retarded and advanced  adiabatic and non-adiabatic Green's functions are given by 
\begin{equation}
G^{R/A} = \Big( \omega - h - \widetilde{\Sigma}^{R/A} - \Sigma_{C}^{(0)} \Big)^{-1},
\end{equation}
\begin{equation}
\widetilde{G}_{(1)}^{R/A} = \Sigma_{C}^{(1)} \Big( G^{R/A} \Big)^2
\end{equation} 
and  
\begin{equation}
\widetilde{G}_{(2)}^{R/A} = \Sigma_{C}^{(1)} G^{R/A} \widetilde{G}_{(1)}^{R/A} + \Sigma_{C}^{(2)} \Big( G^{R/A} \Big)^{2} - \frac{1}{4} \mathcal{C}^{(2)} \Big( G^{R/A} \Big)^4.
\end{equation} 
The adiabatic and non-adiabatic lesser Green's function components are given by
\begin{equation}
G^{<} = G^{R} \widetilde{\Sigma}^{<} G^{A},
\end{equation}
\begin{equation}
\widetilde{G}_{(1)}^{<} = \Sigma_C^{(1)} G^{<} \Big( G^{R} + G^{A} \Big) + \frac{1}{2i} \mathcal{B}^{(1)} G^{R} \partial_{\omega} \widetilde{\Sigma}^{<} G^{A} \Big( G^{R} - G^{A} \Big)
\end{equation}
and 
\begin{multline}
\widetilde{G}_{(2)}^{<} =  G^{R} \Sigma_{C}^{(1)} \widetilde{G}_{(1)}^{<} + G^{R} \Sigma_{C}^{(2)} G^{<} + G^{R} \widetilde{\Sigma}^{<} \widetilde{G}_{(2)}^{A} + \frac{1}{2i} \mathcal{B}^{(1)} \partial_{\omega} \widetilde{\Sigma}^{<} \Big( \widetilde{G}_{(1)}^{R} G^{A} - G^{R} \widetilde{G}_{(1)}^{A} \Big) \\ + \frac{1}{2i} G^{<} \partial_{T} \Sigma^{(1)}_{C} \Big( G^{R} - G^{A} \Big)  - \frac{1}{4} G^{R} G^{A} \Big[ \mathcal{C} \partial_{\omega} \widetilde{\Sigma}^{<} + \Big( \mathcal{B}^{(1)} \Big)^2 \partial^2_{\omega} \widetilde{\Sigma}^{<} \Big] \Big( G^{R} - G^{A} \Big) \\ + \frac{1}{2i} \mathcal{B}^{(2)} G^{R} \partial_{\omega} G^{<} - \frac{1}{8} \mathcal{C} G^{R} \partial^2_{\omega} \widetilde{G}^{<} - \frac{1}{2i} G^{R} \partial_{\omega} \widetilde{\Sigma}^{<} \partial_{T} \widetilde{G}_{(1)}^{A} - \frac{1}{8} G^{R} \partial^2_{\omega} \widetilde{\Sigma}^{<} \partial^2_{T} G^{A}.
\end{multline}
It follows that one can now compute explicit expressions for the perturbative orders of the correlation self-energy and frequently appearing quantities $\mathcal{A}$, $\mathcal{B}$ and $\mathcal{C}$ in section \ref{section3.1}. In the interest of presentation, however, we will not explicitly present expressions for these quantities due to the size of expressions. 

\begin{figure}[t!]
\begin{center}
\includegraphics[width=1.0\columnwidth]{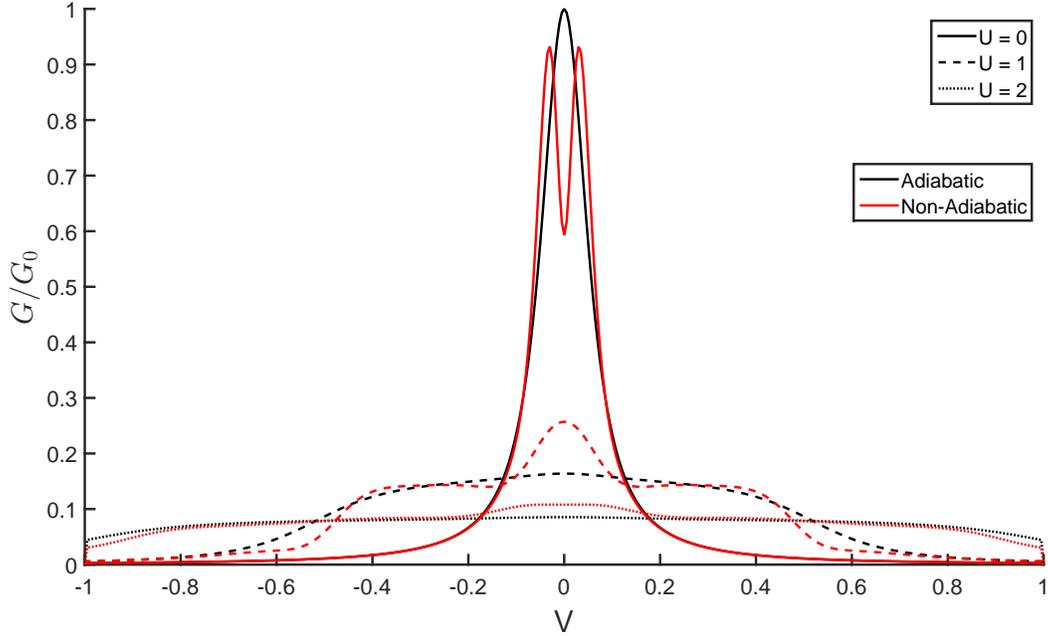}
\end{center}
	\caption{Adiabatic and non-adiabatic differential conductance $G=dJ/dV$  as a function of the applied voltage computed for different values of electron-electron repulsion $U$. The values of the conductance  is given in terms of  $G_0= e^2/h $. Parameters used in calculations: 
$\Gamma_L = \Gamma_R = 0.05$ eV, $\langle \dot{x}^2\rangle =  0.01$ a.u., $\epsilon=0$ eV, and $\Lambda= \Phi = 0.1$ a.u.}
\label{figure2}
\end{figure}

\begin{figure}[t!]
\begin{center}
\includegraphics[width=1.0\columnwidth]{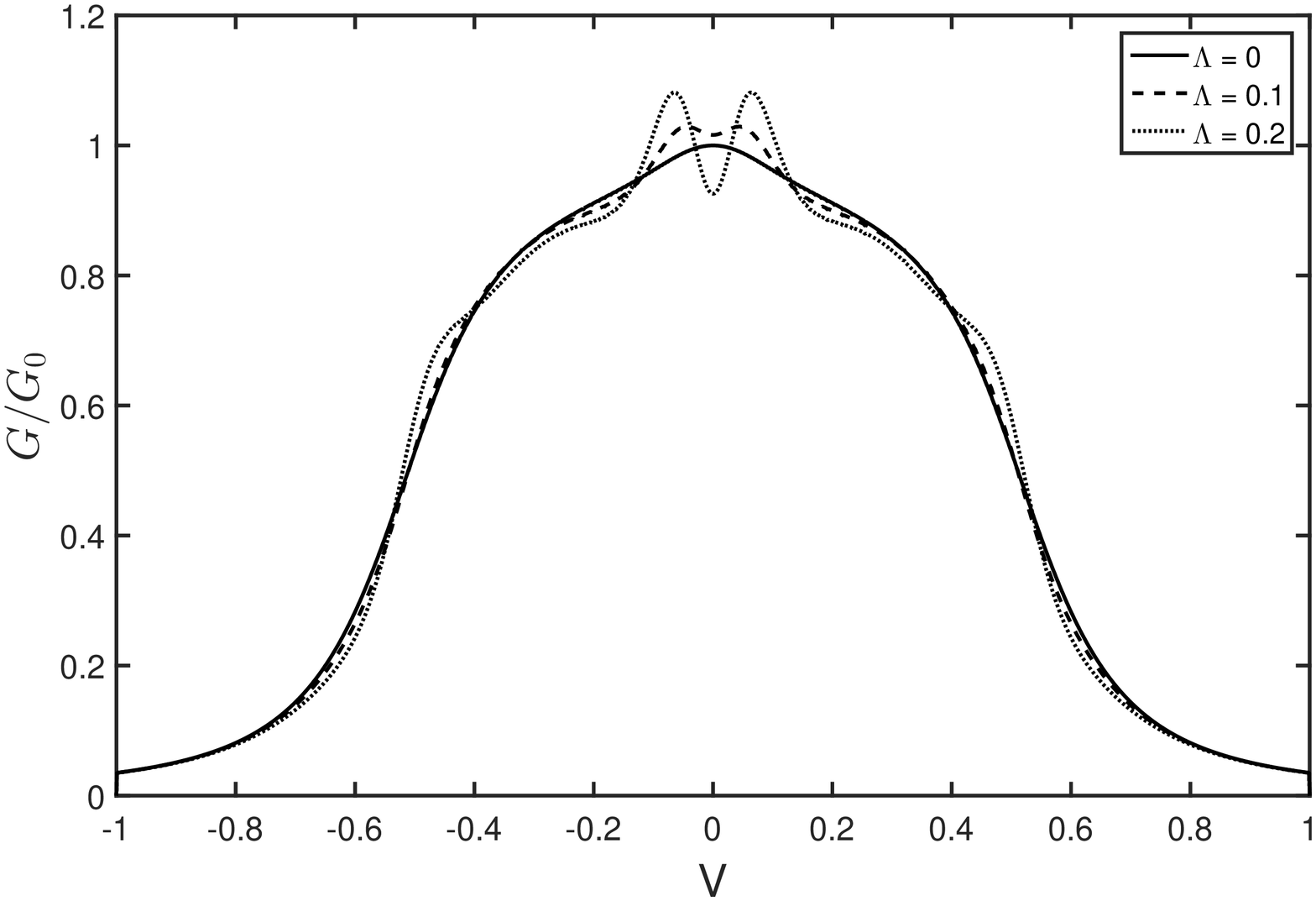}
\end{center}
	\caption{Adiabatic and non-adiabatic differential conductance $G=dJ/dV$   as a function of the applied voltage computed for different values of $\Lambda$. The values of the conductance  is given in terms of  $G_0= e^2/h $. Parameters used in calculations: 
$\Gamma_L = \Gamma_R = 0.05$ eV, $\langle \dot{x}^2\rangle =  4 \times 10^{-4}$ a.u., $\epsilon=0$ eV, $\Phi = 0.1$ a.u. and $U = 0.1$ eV.}
\label{figure2}
\end{figure}

For our model the equations for the adiabatic and second order non-adiabatic current averaged over nuclear velocities are given by (note the we neglect the first order current as it is linear in nuclear velocities and will vanish) 

\begin{equation}
\langle J(t) \rangle = \int d \omega \langle \mathcal{T}(\omega) \rangle \big( f_L - f_R \big),
\label{current}
\end{equation}
where we define the averaged transmission coefficient $ \mathcal{T}(\omega) $ as
\begin{equation}
\langle \mathcal{T}(\omega) \rangle = - 2_s \frac{\Gamma_L \Gamma_R}{\pi \Gamma} \text{Im} \Big\{ G^R + \langle \widetilde{G}_{(2)}^R \rangle - \frac{1}{2i} \langle \partial_{T \omega} \widetilde{G}_{(1)}^R \rangle - \frac{1}{8} \langle \partial_{T \omega}^2 G^R \rangle \Big\}.
\label{transmission}
\end{equation}
Here have used and will continue to use bracket notation (the symbols '$\langle$' and '$\rangle$') to denote velocity averaged quantities (notice that quantities that are not functions of nuclear velocity are not subject to this notation).
In (\ref{transmission}) we see the presence of the factor $2_s$ to represent the spin degeneracy. 

The quantity $ \mathcal{T}(\omega) $ will serve as the main object of study for this section where, in the interest of presentation, we leave the expression for $ \mathcal{T}(\omega) $ as (\ref{transmission}) (representing $ \mathcal{T}(\omega) $ explicitly results in large expressions that are cumbersome and provides little insight). Note that the transmission coefficient is still dependent on all orders of the lesser Green's functions as the retarded components themselves have dependencies on electronic occupation numbers.

Let us now use (\ref{transmission}) to explicitly calculate the  differential conductance  $dJ/dV$ with non-adiabatic contributions for different model parameters. Figure 1 plots these contributions for different electron-electron repulsion strengths $U = 0, 1$ and $2$ eV. For all  values of electron-electron interaction we see the non-adiabatic corrections are mostly pronounced in the resonant transport regime with the conductance decreasing with increasing $U$. In the case of $U = 0$ (corresponding to the absence of electron-electron interaction) the nuclear motion plays destructive role at resonance but slightly increases conductance in the off-resonance regimes, a result that agrees with calculation in our previous work.\cite{kershaw17,kershaw18} Contrary to the non-interacting case, 
non-adiabatic effects contribute constructively  at resonance in the presence of electron-electron interactions in the system. Finally, we observe that the conduction profile width becomes wider with increasing values of $U$. 

Figure 2 considers the differential conductance profile for a specific correlation strength of $U = 0.1$ eV, but instead varies the parameter $\Lambda$ for values $\Lambda = 0, 0.1$ and $0.2$ a.u. We see that the non-adiabatic effects for $\Lambda = 0.1$ a.u. manifest to increase the conductance at and near the resonant situation with regions of destructive and constructive contributions as we move to higher voltages. Selecting $\Lambda = 0.2$ a.u. accentuates the peaks and produces destructive contribution to the molecular conductivity at resonance.

\section{Conclusions}

In this paper, we have developed  a quantum transport theory for interacting electrons which takes into account non-adiabatic effects of nuclear motion. Our approach was based on non-equilibrium Green's functions and the use of Wigner representation to solve the Kadanoff-Baym equations. Slow nuclear motion implies that Green's functions vary slowly with the central time  and oscillate fast with the relative time, with the same argument being applied to the correlation self-energy as well. The time derivatives with respect to central time are used as a small parameter and systematic perturbative expansion is developed to solve the Kadanoff-Baym equations of motion for the Green's functions in the Wigner space. We produced analytic expressions for non-adiabatic electronic Green's functions which depend not solely on instantaneous molecular geometry but likewise on nuclear velocities and accelerations. The general expression for the electric current in terms of Green's functions and self-energies was converted to the Wigner space maintaining terms up to the second order in the central time derivatives. 
As a result, we obtained the formula for electric current through correlated central region with non-adiabatic corrections for time-varying geometry.
Our method allows the systematic treatment of electron-electron interactions and simultaneously includes dynamical effects of nuclear motion.
This theory is concisely illustrated by the calculations of electron transport through the molecular junction described by the Anderson model with dynamically changing single-particle energy level.

\newpage
\appendix
\section{Retarded/Advanced Greens Function Components}
Below is the expression for $\widetilde{G}^{R/A}_{(2)}$ which takes the form
 \begin{multline}
\widetilde{G}^{R/A}_{(2)}  = G^{R/A} \widetilde{\Sigma}^{R/A}_{tot(1)} \widetilde{G}^{R/A}_{(1)} + G^{R/A} \widetilde{\Sigma}^{R/A}_{tot(2)} G^{R/A} + \frac{1}{2i} G^{R/A} \Big( \mathcal{A}_{(1)}^{R/A} G^{R/A} \mathcal{B}^{R/A}_{(1)} G^{R/A} \\ - \mathcal{B}^{R/A}_{(2)} G^{R/A} \mathcal{A}^{R/A}_{(0)} G^{R/A} \Big) - \frac{1}{4} G^{R/A} \Big[ \mathcal{C}^{R/A} G^{R/A} \Big( \mathcal{A}^{R/A}_{(0)} G^{R/A} \Big)^2 + \partial_{T \omega} \widetilde{\Sigma}^{R/A}_{tot(0)} G^{R/A} \\ \times \Big[ \mathcal{B}^{R/A}_{(1)} G^{R/A}, \mathcal{A}^{R/A}_{(0)} G^{R/A} \Big]_{+} + \partial^2_{\omega} \widetilde{\Sigma}^{R/A}_{tot(0)} G^{R/A} \Big( \mathcal{B}^{R/A}_{(1)} G^{R/A} \Big)^2 \Big] - \frac{1}{8} G^{R/A} \\ \times \Big[ \mathcal{C}^{R/A} G^{R/A} \partial^{2}_{\omega} \widetilde{\Sigma}^{R/A}_{tot(0)} G^{R/A} - 2 \Big( \partial_{T \omega} \widetilde{\Sigma}^{R/A}_{tot(0)} G^{R/A} \Big)^2 + \partial^2_{\omega} \widetilde{\Sigma}^{R/A}_{tot(0)} G^{R/A} \partial^{2}_{T} \widetilde{\Sigma}^{R/A}_{T(0)} G^{R/A} \Big] \\ + \frac{1}{2i} G^{R/A} \Big[ \mathcal{A}_{(0)}^{R/A} G^{R/A}, \mathcal{B}^{R/A}_{(1)} G^{R/A} \Big]_{-} \widetilde{\Sigma}^{R/A}_{tot(1)} G^{R/A} + \frac{1}{2i} G^{R/A} \mathcal{B}^{R/A}_{(1)} G^{R/A} \partial_{\omega} \widetilde{\Sigma}^{R/A}_{tot(1)} G^{R/A} \\ + \frac{1}{2i} G^{R/A} \mathcal{A}_{(0)} G^{R/A} \partial_{T} \widetilde{\Sigma}^{R/A}_{tot(1)} G^{R/A} - \frac{1}{2i} G^{R/A} \mathcal{B}^{R/A}_{(1)} G^{R/A} \widetilde{\Sigma}^{R/A}_{tot(1)} G^{R/A} \mathcal{A}^{R/A}_{(0)} G^{R/A} \\ + \frac{1}{2i} G^{R/A} \mathcal{A}_{(0)} G^{R/A} \widetilde{\Sigma}^{R/A}_{tot(1)} G^{R/A} \mathcal{B}^{R/A}_{(1)} G^{R/A} - \frac{1}{4} G^{R/A} \Big[ \mathcal{A}_{(0)}^{R/A} G^{R/A}, \mathcal{B}^{R/A}_{(1)} G^{R/A} \Big]_{-} \\ \times \Big[ \mathcal{A}_{(0)}^{R/A} G^{R/A}, \mathcal{B}^{R/A}_{(1)} G^{R/A} \Big]_{-} + \frac{1}{2i} \mathcal{B}^{R/A}_{(1)} G^{R/A} \Big[ \mathcal{B}^{R/A}_{(1)} G^{R/A}, \partial^2_{\omega} \widetilde{\Sigma}^{R/A}_{tot(0)} G^{R/A} \Big]_{-} + \frac{1}{2i} \mathcal{A}_{(0)}^{R/A} \\ \times G^{R/A} \Big[ \mathcal{B}^{R/A}_{(1)} G^{R/A}, \partial_{T \omega} \widetilde{\Sigma}^{R/A}_{tot(0)} G^{R/A} \Big]_{-} + \frac{1}{2i} \mathcal{B}^{R/A}_{(1)} G^{R/A} \Big[ \mathcal{B}^{R/A}_{(1)} G^{R/A}, \Big( \mathcal{A}_{(0)}^{R/A} G^{R/A} \Big)^2 \Big]_{-} \\ + \frac{1}{2i} \mathcal{A}_{(0)}^{R/A} G^{R/A} \Big[ \mathcal{A}_{(0)}^{R/A} G^{R/A} \mathcal{B}^{R/A}_{(1)} G^{R/A}, \mathcal{B}^{R/A}_{(1)} G^{R/A} \Big]_{-} + \frac{1}{2i} \mathcal{B}^{R/A}_{(1)} G^{R/A} \\ \times \Big[ \mathcal{A}_{(0)}^{R/A} G^{R/A}, \partial_{T \omega} \widetilde{\Sigma}^{R/A}_{tot(0)} G^{R/A} \Big]_{-} + \frac{1}{2i} \mathcal{A}_{(0)}^{R/A} G^{R/A} \Big[ \mathcal{A}_{(0)}^{R/A} G^{R/A}, \mathcal{C}^{R/A} G^{R/A} \Big]_{-} \\ + \frac{1}{2i} \mathcal{B}^{R/A}_{(1)} G^{R/A} \Big[ \mathcal{B}^{R/A}_{(1)} G^{R/A} \mathcal{A}_{(0)}^{R/A} G^{R/A}, \mathcal{A}_{(0)}^{R/A} G^{R/A} \Big]_{-} + \frac{1}{2i} \mathcal{A}_{(0)}^{R/A} G^{R/A} \\ \times \Big[ \mathcal{A}_{(0)}^{R/A} G^{R/A}, \Big( \mathcal{B}^{R/A}_{(1)} G^{R/A} \Big)^2 \Big]_{-}.
\end{multline}
where, in the expression above, we have chosen to keep expressions for $\widetilde{G}^{R/A}_{(1)}$ as an input.

\section{Computing Green's Function Derivatives}
\subsection{Computing $G^{R} \hat{D}^{R}_{a} G^{<}$ Term}
\label{Appendix A1}
We now calculate the quantity $G^{R} \hat{D}^{R}_{a} G^{<}$. We know that $G^{R} \hat{D}^{R}_{a} G^{<}$ has the explicit form 
\begin{equation}
G^{R} \hat{D}^{R}_{a} G^{<} = - \frac{1}{8} G^{R} \Big( \mathcal{C}^{R} \partial^{2}_{\omega} - 2 \partial_{T \omega} \widetilde{\Sigma}^{R}_{tot(0)} \partial_{T \omega} + \partial^{2}_{\omega} \widetilde{\Sigma}^{R}_{tot(0)} \partial^{2}_{T} \Big) G^{<}
\end{equation}
which becomes 
\begin{equation}
G^{R} \hat{D}^{R}_{a} G^{<} = - \frac{1}{8} G^{R} \Big( \mathcal{C}^{R} \partial^{2}_{\omega} G^{<} - 2 \partial_{T \omega} \widetilde{\Sigma}^{R}_{tot(0)} \partial_{T \omega} G^{<} + \partial^{2}_{\omega} \widetilde{\Sigma}^{R}_{tot(0)} \partial^{2}_{T} G^{<} \Big).
\end{equation}
We take note that
\begin{multline}
\partial^{2}_{\omega} G^{<} = 2 \Big( G^{R} \mathcal{A}^{R}_{(0)} \Big)^2 G^{<} + 2 G^{<} \Big( \mathcal{A}^{A}_{(0)} G^{A} \Big)^2 - 2 G^{R} \mathcal{A}^{R}_{(0)} G^{R} \partial_{\omega} \widetilde{\Sigma}^{<}_{tot(0)} G^{A} \\[0.5ex] + 2 G^{R} \mathcal{A}^{R}_{(0)} G^{<} \mathcal{A}^{A}_{(0)} G^{A} - 2 G^{R} \partial_{\omega} \widetilde{\Sigma}^{<}_{tot(0)} G^{A} \mathcal{A}^{A}_{(0)} G^{A} + G^{R} \partial_{\omega}^{2} \widetilde{\Sigma}^{R}_{tot(0)} G^{<} + G^{R} \partial_{\omega}^{2} \widetilde{\Sigma}^{<}_{tot(0)} G^{A} \\[0.5ex] + G^{<} \partial_{\omega}^{2} \widetilde{\Sigma}^{A}_{tot(0)} G^{A},
\end{multline}
\begin{multline}
\partial_{T \omega} G^{<} =  - G^{R} \Big[ \mathcal{B}^{R}_{(1)} G^{R}, \mathcal{A}^{R}_{(0)} G^{R} \Big]_{+} \widetilde{\Sigma}^{<}_{tot(0)} G^{A} - G^{R} \mathcal{A}^{R}_{(0)} G^{<} \mathcal{B}^{A}_{(1)} G^{A}  - G^{R} \mathcal{B}^{R}_{(1)} G^{<} \mathcal{A}^{A}_{(0)} G^{A} \\[0.5ex] - G^{<} \Big[ \mathcal{B}^{A}_{(1)} G^{A}, \mathcal{A}^{A}_{(0)} G^{A} \Big]_{+} - G^{R} \mathcal{A}^{R}_{(0)} G^{R} \partial_{T} \widetilde{\Sigma}^{<}_{tot(0)} G^{A} - G^{R} \partial_{T} \widetilde{\Sigma}^{<}_{tot(0)} G^{A}  \mathcal{A}^{A}_{(0)} G^{A} \\[0.5ex] + G^{R} \mathcal{B}^{R}_{(1)} G^{R} \partial_{\omega} \widetilde{\Sigma}^{<}_{tot(0)} G^{A} + G^{R} \partial_{\omega} \widetilde{\Sigma}^{<}_{tot(0)} G^{A} \mathcal{B}^{A}_{(1)} G^{A} + G^{<} \partial_{T \omega} \widetilde{\Sigma}^{A}_{tot(0)} G^{A} + G^{R} \partial_{T \omega} \widetilde{\Sigma}^{R}_{tot(0)} G^{<} \\[0.5ex] + G^{R} \partial_{T \omega} \widetilde{\Sigma}^{<}_{tot(0)} G^{A} 
\end{multline}
and 
\begin{multline}
\partial^{2}_{T} G^{<} =  2 \Big( G^{R} \mathcal{B}^{R}_{(1)} \Big)^2 G^{<} + 2 G^{<} \Big( \mathcal{B}^{A}_{(1)} G^{A} \Big)^2 + 2 G^{R} \mathcal{B}^{R}_{(1)} G^{R} \partial_{T} \widetilde{\Sigma}^{<}_{tot(0)} G^{A} \\[0.5ex] + 2 G^{R} \mathcal{B}^{R}_{(1)} G^{<} \mathcal{B}^{A}_{(1)} G^{A} + 2 G^{R} \partial_{T} \widetilde{\Sigma}^{<}_{tot(0)} G^{A} \mathcal{B}^{A}_{(1)} G^{A} + G^{R} \mathcal{C}^{A} G^{<} + G^{R} \partial_{T}^{2} \widetilde{\Sigma}^{<}_{tot(0)} G^{A} \\[0.5ex] + G^{<} \mathcal{C}^{R} G^{A}.
\end{multline}
Substituting in the above expressions we conclude that 
\begin{multline}
G^{R} \hat{D}^{R}_{a} G^{<} = - \frac{1}{8} G^{R} \mathcal{C}^{R} \Big[ 2 \Big( G^{R} \mathcal{A}^{R}_{(0)} \Big)^2 G^{<} + 2 G^{<} \Big( \mathcal{A}^{A}_{(0)} G^{A} \Big)^2 - 2 G^{R} \mathcal{A}^{R}_{(0)} G^{R} \partial_{\omega} \widetilde{\Sigma}^{<}_{tot(0)} \\ \times G^{A} + 2 G^{R} \mathcal{A}^{R}_{(0)} G^{<} \mathcal{A}^{A}_{(0)} G^{A} - 2 G^{R} \partial_{\omega} \widetilde{\Sigma}^{<}_{tot(0)} G^{A} \mathcal{A}^{A}_{(0)} G^{A} + G^{R} \partial_{\omega}^{2} \widetilde{\Sigma}^{R}_{tot(0)} G^{<} + G^{R} \partial_{\omega}^{2} \widetilde{\Sigma}^{<}_{tot(0)} \\ \times G^{A} + G^{<} \partial_{\omega}^{2} \widetilde{\Sigma}^{A}_{tot(0)} G^{A} \Big] + \frac{1}{4} G^{R} \partial_{T \omega} \widetilde{\Sigma}^{R}_{tot(0)} \Big(  - G^{R} \Big[ \mathcal{B}^{R}_{\alpha} G^{R}, \mathcal{A}^{R}_{(0)} G^{R} \Big]_{+} \widetilde{\Sigma}^{<}_{tot(0)} G^{A} \\ - G^{R} \mathcal{A}^{R}_{(0)} G^{<} \mathcal{B}^{A}_{(1)} G^{A} - G^{R} \mathcal{B}^{R}_{(1)} G^{<} \mathcal{A}^{A}_{(0)} G^{A} - G^{<} \Big[ \mathcal{B}^{A}_{(1)} G^{A}, \mathcal{A}^{A}_{(0)} G^{A} \Big]_{+} \\ - G^{R} \mathcal{A}^{R}_{(0)} G^{R} \partial_{T} \widetilde{\Sigma}^{<}_{tot(0)} G^{A} - G^{R} \partial_{T} \widetilde{\Sigma}^{<}_{tot(0)} G^{A} \mathcal{A}^{A}_{(0)} G^{A} + G^{R} \mathcal{B}^{R}_{(1)} G^{R} \partial_{\omega} \widetilde{\Sigma}^{<}_{tot(0)} G^{A} \\[0.5ex] + G^{R} \partial_{\omega} \widetilde{\Sigma}^{<}_{tot(0)} G^{A} \mathcal{B}^{A}_{(1)} G^{A} + G^{<} \partial_{T \omega} \widetilde{\Sigma}^{A}_{tot(0)} G^{A} + G^{R} \partial_{T \omega} \widetilde{\Sigma}^{R}_{tot(0)} G^{<} + G^{R} \partial_{T \omega} \widetilde{\Sigma}^{<}_{tot(0)} G^{A} \Big) \\ - \frac{1}{8} G^{R} \partial^{2}_{\omega} \widetilde{\Sigma}^{R}_{tot(0)} \Big[ 2 \Big( G^{R} \mathcal{B}^{R}_{(1)} \Big)^2 G^{<} + 2 G^{<} \Big( \mathcal{B}^{A}_{(1)} G^{A} \Big)^2 + 2 G^{R} \mathcal{B}^{R}_{(1)} G^{R} \partial_{T} \widetilde{\Sigma}^{<}_{tot(0)} G^{A} \\ + 2 G^{R} \mathcal{B}^{R}_{(1)} G^{<} \mathcal{B}^{A}_{(1)} G^{A} + 2 G^{R} \partial_{T} \widetilde{\Sigma}^{<}_{tot(0)} G^{A} \mathcal{B}^{A}_{(1)} G^{A} + G^{R} \mathcal{C}^{A} G^{<} + G^{R} \partial_{T}^{2} \widetilde{\Sigma}^{<}_{tot(0)} G^{A} \\[0.5ex] + G^{<} \mathcal{C}^{R} G^{A} \Big],
\end{multline}
which concludes the derivation of $G^{R} \hat{D}^{R}_{a} G^{<}$. 

\subsection{Computing $G^{R/A} \hat{P} \widetilde{G}_{(1)}^{R/A}$ Term}
\label{Appendix A2}
We compute $G^{R/A} \hat{P}$ acting on $\widetilde{G}_{(1)}^{R/A}$ in the case of arbitrary matrix coefficients $X$ and $Y$. We know that $G^{R/A} \hat{P} \widetilde{G}_{(1)}^{R/A}$ has the explicit form 
\begin{equation}
G^{R/A} \hat{P} \widetilde{G}_{(1)}^{R/A} = \frac{1}{2i} G^{R/A} \Big(  X \partial_{\omega} +  Y \partial_{T} \Big) \widetilde{G}_{(1)}^{R/A}.
\end{equation}
We know from a previous section that
\begin{equation}
\widetilde{G}^{R/A}_{(1)} = G^{R/A} \widetilde{\Sigma}^{R/A}_{tot(1)} G^{R/A} + \frac{1}{2i} G^{R/A} \Big[ \mathcal{A}_{(0)}^{R/A} G^{R/A}, \mathcal{B}^{R/A}_{(1)} G^{R/A} \Big]_{-}.
\end{equation}  
Thus, it is found that $\hat{P} \widetilde{G}_{(1)}^{R/A}$ becomes
\begin{multline}
\hat{P} \widetilde{G}^{R/A}_{(1)} = \frac{1}{2i} G^{R/A} \Big(  X \partial_{\omega} +  Y \partial_{T} \Big) G^{R/A} \widetilde{\Sigma}^{R/A}_{tot(1)} G^{R/A} \\ - \frac{1}{4} G^{R/A} \Big(  X \partial_{\omega} +  Y \partial_{T} \Big) G^{R/A} \Big[ \mathcal{A}_{(0)}^{R/A} G^{R/A}, \mathcal{B}^{R/A}_{(1)} G^{R/A} \Big]_{-}.
\end{multline} 
We note that 
\begin{multline}
G^{R/A} \hat{P} \widetilde{G}^{R/A}_{(1)} = \frac{1}{2i} G^{R/A} \Big( Y G^{R/A} \mathcal{B}^{R/A}_{(1)} G^{R/A} - X G^{R/A} \mathcal{A}^{R/A}_{(0)} G^{R/A} \Big) \widetilde{\Sigma}^{R/A}_{tot(1)} G^{R/A} \\ + \frac{1}{2i} G^{R/A} X G^{R/A} \partial_{\omega} \widetilde{\Sigma}^{R/A}_{tot(1)} G^{R/A} + \frac{1}{2i} G^{R/A} Y G^{R/A} \partial_{T} \widetilde{\Sigma}^{R/A}_{tot(1)} G^{R/A} \\ - \frac{1}{2i} G^{R/A} X G^{R/A} \widetilde{\Sigma}^{R/A}_{tot(1)} G^{R/A} \mathcal{A}^{R/A}_{(0)} G^{R/A} + \frac{1}{2i} G^{R/A} Y G^{R/A} \widetilde{\Sigma}^{R/A}_{tot(1)} G^{R/A} \mathcal{B}^{R/A}_{(1)} G^{R/A} \\ - \frac{1}{4} G^{R/A} \Big( Y G^{R/A} \mathcal{B}^{R/A}_{(1)} G^{R/A} - X G^{R/A} \mathcal{A}^{R/A}_{(0)} G^{R/A} \Big) \Big[ \mathcal{A}_{(0)}^{R/A} G^{R/A}, \mathcal{B}^{R/A}_{(1)} G^{R/A} \Big]_{-} \\ + \frac{1}{2i} G^{R/A} X G^{R/A} \Big[ \mathcal{B}^{R/A}_{(1)} G^{R/A}, \partial^2_{\omega} \widetilde{\Sigma}^{R/A}_{tot(0)} G^{R/A} \Big]_{-} + \frac{1}{2i} G^{R/A} Y G^{R/A} \\ \times \Big[ \mathcal{B}^{R/A}_{(1)} G^{R/A}, \partial_{T \omega} \widetilde{\Sigma}^{R/A}_{tot(0)} G^{R/A} \Big]_{-} + \frac{1}{2i} G^{R/A} X G^{R/A} \Big[ \mathcal{B}^{R/A}_{(1)} G^{R/A}, \Big( \mathcal{A}_{(0)}^{R/A} G^{R/A} \Big)^2 \Big]_{-} \\ + \frac{1}{2i} G^{R/A} Y G^{R/A} \Big[ \mathcal{A}_{(0)}^{R/A} G^{R/A} \mathcal{B}^{R/A}_{(1)} G^{R/A}, \mathcal{B}^{R/A}_{(1)} G^{R/A} \Big]_{-} + \frac{1}{2i} G^{R/A} X G^{R/A} \\ \times \Big[ \mathcal{A}_{(0)}^{R/A} G^{R/A}, \partial_{T \omega} \widetilde{\Sigma}^{R/A}_{tot(0)} G^{R/A} \Big]_{-} + \frac{1}{2i} G^{R/A} Y G^{R/A} \Big[ \mathcal{A}_{(0)}^{R/A} G^{R/A}, \mathcal{C}^{R/A} G^{R/A} \Big]_{-} \\ + \frac{1}{2i} G^{R/A} X G^{R/A} \Big[ \mathcal{B}^{R/A}_{(1)} G^{R/A} \mathcal{A}_{(0)}^{R/A} G^{R/A}, \mathcal{A}_{(0)}^{R/A} G^{R/A} \Big]_{-} + \frac{1}{2i} G^{R/A} Y G^{R/A} \\ \times \Big[ \mathcal{A}_{(0)}^{R/A} G^{R/A}, \Big( \mathcal{B}^{R/A}_{(1)} G^{R/A} \Big)^2 \Big]_{-},
\label{referee}
\end{multline} 
which concludes the derivation of $\hat{P} \widetilde{G}^{R/A}_{(1)}$. The above expression is altered for the relevant matrix quantities to get $\hat{P}_{a(1)}^{R/A} \widetilde{G}^{R/A}_{(1)}$ and $\hat{P}_{b(1)}^{<} \widetilde{G}^{A}_{(1)}$ required for the solution of the retarded/advanced equations of motion and the lesser equations of motion respectively. When considering $\hat{P}_{a(1)}^{R/A} \widetilde{G}^{R/A}_{(1)}$ in section \ref{RAsection}, one chooses $X = \mathcal{B}^{R/A}_{(1)}$ and $Y = \mathcal{A}^{R/A}_{(0)}$ in (\ref{referee}) where this choice of $X$ and $Y$ allows one use commutator notation to simplify the expressions. When considering $\hat{P}_{a(2)}^{<} \widetilde{G}^{A}_{(1)}$ then it follows that one chooses $X = \partial_{T} \widetilde{\Sigma}^{<}_{tot(0)}$ and $Y = - \partial_{\omega} \widetilde{\Sigma}^{<}_{tot(0)}$ in (\ref{referee}) and select the appropriate retarded and advanced components.

Note that Appendix A3 makes use of this derivation as well (clearly labeled) where one selects $X = \mathcal{B}^{R/A}_{(1)} G^{R} \widetilde{\Sigma}^{<}_{tot(0)}$ and $Y = \mathcal{A}^{R/A}_{(0)} G^{R} \widetilde{\Sigma}^{<}_{tot(0)}$. 

\subsection{Computing $G^{R} \hat{P}^{R}_{a(1)} \widetilde{G}_{(1)}^{<}$ Term}
\label{Appendix A3}
We compute $G^{R} \hat{P}^{R}_{a}$ acting on $\widetilde{G}_{(1)}^{<}$. We know that $G^{R} \hat{P}^{R}_{a} \widetilde{G}_{(1)}^{<}$ has the explicit form 
\begin{equation}
G^{R} \hat{P}^{R}_{a(1)} \widetilde{G}_{(1)}^{<} = \frac{1}{2i} G^{R} \Big( \mathcal{B}^{R}_{(1)} \partial_{\omega} +  \mathcal{A}^{R}_{(0)} \partial_{T} \Big) \widetilde{G}_{(1)}^{<}.
\end{equation}
From a previous calculation it is known that 
\begin{multline}
\widetilde{G}^{<}_{(1)} = G^{R} \widetilde{\Sigma}^{R}_{tot(1)} G^{<} + G^{R} \widetilde{\Sigma}^{<}_{tot(0)} \widetilde{G}^{A}_{(1)} + G^{R} \widetilde{\Sigma}^{<}_{tot(1)} G^{A} + \frac{1}{2i} G^{R} \Big( \mathcal{A}_{(0)}^{R} G^{R} \mathcal{B}^{R}_{(1)} \widetilde{G}^{<} \\ + \mathcal{A}_{(0)}^{R} G^{<} \mathcal{B}^{A}_{(1)} G^{A} - \mathcal{B}_{(1)}^{R} G^{R} \mathcal{A}^{R}_{(0)} G^{<} - \mathcal{B}_{(1)}^{R} G^{<} \mathcal{A}^{A}_{(0)} G^{A} \Big) \\ + \frac{1}{2i} G^{R} \Big( \mathcal{B}_{(1)}^{R} G^{R} \partial_{\omega} \widetilde{\Sigma}^{<}_{tot(0)} G^{A} + \mathcal{A}_{(0)}^{R} G^{R} \partial_{T} \widetilde{\Sigma}^{<}_{tot(0)} G^{A} \Big) \\ - \frac{1}{2i} G^{R} \Big( \partial_{\omega} \widetilde{\Sigma}^{<}_{tot(0)} G^{A} \mathcal{B}^{A}_{(1)} G^{A} + \partial_{T} \widetilde{\Sigma}^{<}_{tot(0)} G^{A} \mathcal{A}^{A}_{(0)} G^{A} \Big) .
\end{multline}
We break the expression into four components such that 
\begin{equation}
A_1 = G^{R} \widetilde{\Sigma}^{R}_{tot(1)} G^{<} + G^{R} \widetilde{\Sigma}^{<}_{tot(0)} \widetilde{G}^{A}_{(1)} + G^{R} \widetilde{\Sigma}^{<}_{tot(1)} G^{A},
\end{equation}
\begin{equation}
A_2 = \frac{1}{2i} G^{R} \Big( \mathcal{A}_{(0)}^{R} G^{R} \mathcal{B}^{R}_{(1)} G^{<} + \mathcal{A}_{(0)}^{R} G^{<} \mathcal{B}^{A}_{(1)} G^{A} \\ - \mathcal{B}_{(1)}^{R} G^{R} \mathcal{A}^{R}_{(0)} G^{<} - \mathcal{B}_{(1)}^{R} G^{<}  \mathcal{A}^{A}_{(0)} G^{A} \Big),
\end{equation}
\begin{equation}
A_3 = \frac{1}{2i} G^{R} \Big( \mathcal{B}_{(1)}^{R} G^{R} \partial_{\omega} \widetilde{\Sigma}^{<}_{tot(0)} G^{A} + \mathcal{A}_{(0)}^{R} G^{R} \partial_{T} \widetilde{\Sigma}^{<}_{tot(0)} G^{A} \Big)
\end{equation}
and
\begin{equation}
A_4 = - \frac{1}{2i} G^{R} \Big( \partial_{\omega} \widetilde{\Sigma}^{<}_{tot(0)} G^{A} \mathcal{B}^{A}_{(1)} G^{A} + \partial_{T} \widetilde{\Sigma}^{<}_{tot(0)} G^{A} \mathcal{A}^{A}_{(0)} G^{A} \Big),
\end{equation}
where it obviously follows that 
\begin{equation}
\widetilde{G}^{<}_{(1)} = A_1 + A_2 + A_3 + A_4. 
\end{equation}
Computing the quantity $G^{R} \hat{P}^{R}_{a} \widetilde{G}_{(1)}^{<}$ is then represented as 
\begin{equation}
G^{R} \hat{P}^{R}_{a(1)} \widetilde{G}^{<}_{(1)} = G^{R} \hat{P}^{R}_{a(1)} A_1 + G^{R} \hat{P}^{R}_{a(1)} A_2 + G^{R} \hat{P}^{R}_{a(1)} A_3 + G^{R} \hat{P}^{R}_{a(1)} A_4. 
\label{combination}
\end{equation}
Presenting $G^{R} \hat{P}^{R}_{a(1)} \widetilde{G}^{<}_{(1)}$ explicitly leads to very large expressions that are difficult to interpret and check. To improve presentation we will present expressions for the components defined above. Through a long process one can show that 
\begin{multline}
{\color{black} G^{R}} \hat{P}^{R}_{a(1)} A_1 = \frac{1}{2i} {\color{black} G^{R}} \Big[ \mathcal{A}^{R}_{(0)} {\color{black} G^{R}}, {\color{black} \mathcal{B}^{R}_{(1)}} {\color{black} G^{R}} \Big]_{-} \widetilde{\Sigma}^{R}_{{\color{black} tot}(1)} {\color{black} G^{<}} + \frac{1}{2i} {\color{black} G^{R}} \Big[ \mathcal{A}^{R}_{(0)} {\color{black} G^{R}}, {\color{black} \mathcal{B}^{R}_{(1)}} {\color{black} G^{R}} \Big]_{-} \widetilde{\Sigma}^{<}_{{\color{black} tot}(0)} \widetilde{G}^{A}_{(1)} \\[0.5ex] + \frac{1}{2i} \widetilde{G}^{R} \Big[ \mathcal{A}^{R}_{(0)} {\color{black} G^{R}}, {\color{black} \mathcal{B}^{R}_{(1)}} {\color{black} G^{R}} \Big]_{-} \widetilde{\Sigma}^{<}_{{\color{black} tot}(1)} {\color{black} G^{A}} + \frac{1}{2i} {\color{black} G^{R}} \Big( {\color{black} \mathcal{B}^{R}_{(1)}} {\color{black} G^{R}} \partial_{\omega} \widetilde{\Sigma}^{R}_{{\color{black} tot}(1)} + \mathcal{A}^{R}_{(0)} {\color{black} G^{R}} \partial_{T} \widetilde{\Sigma}^{R}_{{\color{black} tot}(1)} \Big) \\ \times {\color{black} G^{<}} + \frac{1}{2i} {\color{black} G^{R}} \Big( {\color{black} \mathcal{B}^{R}_{(1)}} {\color{black} G^{R}} \partial_{\omega} \widetilde{\Sigma}^{<}_{{\color{black} tot}(0)} +  \mathcal{A}^{R}_{(0)} {\color{black} G^{R}} \partial_{T} \widetilde{\Sigma}^{<}_{ {\color{black} tot}(0)} \Big) \widetilde{G}^{A}_{(1)} + \frac{1}{2i} {\color{black} G^{R}} \Big( {\color{black} \mathcal{B}^{R}_{(1)}} {\color{black} G^{R}} \partial_{\omega} \widetilde{\Sigma}^{<}_{{\color{black} tot}(1)} \\ + \mathcal{A}^{R}_{(0)} {\color{black} G^{R}} \partial_{T} \widetilde{\Sigma}^{<}_{{\color{black} tot}(1)} \Big) {\color{black} G^{A}} + \frac{1}{2i} {\color{black} G^{R}} \Big[ {\color{black} \mathcal{B}^{R}_{(1)}} {\color{black} G^{R}} \widetilde{\Sigma}^{R}_{{\color{black} tot}(1)} \Big( - {\color{black} G^{R}} \mathcal{A}^{R}_{(0)} {\color{black} G^{<}} - {\color{black} G^{<}}  \mathcal{A}^{A}_{(0)} {\color{black} G^{A}} \\ + {\color{black} G^{R}} \partial_{\omega} \widetilde{\Sigma}^{<}_{{\color{black} tot} (0)} {\color{black} G^{A}} \Big) + \mathcal{A}^{R}_{(0)} {\color{black} G^{R}} \widetilde{\Sigma}^{R}_{{\color{black} tot}(1)} \Big( {\color{black} G^{R}} {\color{black} \mathcal{B}^{R}_{(1)}} {\color{black} G^{<}} + {\color{black} G^{<}} \mathcal{B}^{A}_{\alpha(1)} {\color{black} G^{A}} + {\color{black} G^{R}} \partial_{T} \widetilde{\Sigma}^{<}_{{\color{black} tot} (0)} {\color{black} G^{A}} \Big) \Big] \\ + \underbrace{\frac{1}{2i} {\color{black} G^{R}} \Big( {\color{black} \mathcal{B}^{R}_{(1)}} {\color{black} G^{R}} \widetilde{\Sigma}^{<}_{{\color{black} tot}(0)} \partial_{\omega}  + \mathcal{A}^{R}_{(0)} {\color{black} G^{R}} \widetilde{\Sigma}^{<}_{{\color{black} tot}(0)} \partial_{T} \Big) \widetilde{G}_{(1)}^{A}}_{\text{Appendix \ref{Appendix A2}}} + \frac{1}{2i} {\color{black} G^{R}} \Big( \mathcal{A}^{R}_{(0)} {\color{black} G^{R}} \widetilde{\Sigma}^{<}_{{\color{black} tot}(1)} {\color{black} G^{A}} {\color{black} \mathcal{B}^{A}_{(1)}} \\ \times {\color{black} G^{A}} - {\color{black} \mathcal{B}^{R}_{(1)}} {\color{black} G^{R}} \widetilde{\Sigma}^{<}_{{\color{black} tot}(1)} {\color{black} G^{A}} \mathcal{A}^{R}_{(0)} {\color{black} G^{A}} \Big),
\end{multline}
\begin{multline}
{\color{black} G^{R}} \hat{P}^{R}_{a(1)} A_2 = - \frac{1}{4} {\color{black} G^{R}} \Big[ \mathcal{A}^{R}_{(0)} {\color{black} G^{R}}, {\color{black} \mathcal{B}^{R}_{(1)}} {\color{black} G^{R}} \Big]_{-} \Big( \mathcal{A}_{(0)}^{R} {\color{black} G^{R}} {\color{black} \mathcal{B}^{R}_{(1)}} {\color{black} G^{<}} + \mathcal{A}_{(0)}^{R} {\color{black} G^{<}} {\color{black} \mathcal{B}^{A}_{(1)}} {\color{black} G^{A}} \\ - {\color{black} \mathcal{B}_{(1)}^{R}} {\color{black} G^{R}} \mathcal{A}^{R}_{(0)} {\color{black} G^{<}} - {\color{black} \mathcal{B}_{(1)}^{R}} {\color{black} G^{<}} \mathcal{A}^{A}_{(0)} {\color{black} G^{A}} \Big) + \frac{1}{4} {\color{black} G^{R}} \Big[ \Big( {\color{black} \mathcal{B}^{R}_{(1)}} \partial^2_{\omega} \widetilde{\Sigma}^{R}_{{\color{black}  tot}(0)} + \mathcal{A}^{R}_{(0)} \partial_{T \omega} \widetilde{\Sigma}^{R}_{{\color{black}  tot}(0)} \Big)   \Big( {\color{black} G^{R}} \\ \times {\color{black} \mathcal{B}^{R}_{(1)}} {\color{black} G^{<}} + {\color{black} G^{<}} \mathcal{B}^{A}_{\alpha(1)} {\color{black} G^{A}} \Big) + \Big( {\color{black} \mathcal{B}^{R}_{(1)}} \partial_{T \omega} \widetilde{\Sigma}^{R}_{{\color{black}  tot}(0)} + \mathcal{A}^{R}_{(0)} {\color{black} \mathcal{C}^{R}} \Big) \Big( {\color{black} G^{R}} \mathcal{A}^{R}_{(0)} {\color{black} G^{<}} + {\color{black} G^{<}} \mathcal{A}^{A}_{(0)} \\ \times {\color{black} G^{A}} \Big) \Big] - \frac{1}{4} {\color{black} G^{R}} \Big[ \Big( \mathcal{A}^{R}_{(0)} {\color{black} G^{R}} \Big)^2 , {\color{black} \mathcal{B}^{R}_{(1)}} {\color{black} G^{R}} \Big]_{-} {\color{black} \mathcal{B}^{R}_{(1)}} {\color{black} G^{<}}  + \frac{1}{4} {\color{black} G^{R}} \Big[ \mathcal{A}^{R}_{(0)} {\color{black} G^{R}} , {\color{black} \mathcal{B}^{R}_{(1)}} {\color{black} G^{R}} {\color{black} \mathcal{B}^{R}_{(1)}} {\color{black} G^{R}} \Big]_{-} \\ \times \mathcal{A}^{R}_{(0)} {\color{black} G^{<}} - \frac{1}{4} {\color{black} G^{R}} \Big[ - {\color{black} \mathcal{B}^{R}_{(1)}} \Big( {\color{black} G^{R}} \mathcal{A}^{R}_{(0)} \Big)^2 {\color{black} G^{<}} - {\color{black} \mathcal{B}^{R}_{(1)}} {\color{black} G^{R}} \mathcal{A}^{R}_{(0)} {\color{black} G^{<}} \mathcal{A}^{A}_{(0)} {\color{black} G^{A}} + {\color{black} \mathcal{B}^{R}_{(1)}} {\color{black} G^{R}} \\ \times \mathcal{A}^{R}_{(0)} {\color{black} G^{R}} \partial_{\omega} \widetilde{\Sigma}^{<}_{{\color{black}  tot}(0)} {\color{black} G^{A}} + \Big( \mathcal{A}^{R}_{(0)} {\color{black} G^{R}} \Big)^2 {\color{black} \mathcal{B}^{R}_{(1)}} {\color{black} G^{<}} + \mathcal{A}^{R}_{(0)} {\color{black} G^{R}} \mathcal{A}^{R}_{(0)} {\color{black} G^{<}} {\color{black} \mathcal{B}^{A}_{(1)}} {\color{black} G^{A}} + \Big( \mathcal{A}^{R}_{(0)} {\color{black} G^{R}} \Big)^2 \\ \times \partial_{T} \widetilde{\Sigma}^{<}_{{\color{black}  tot}(0)} {\color{black} G^{A}} \Big] {\color{black} \mathcal{B}^{A}_{(1)}} {\color{black} G^{A}} + \frac{1}{4} {\color{black} G^{R}} \Big[ - {\color{black} \mathcal{B}^{R}_{(1)}} {\color{black} G^{R}} {\color{black} \mathcal{B}^{R}_{(1)}} {\color{black} G^{R}} \mathcal{A}^{R}_{(0)} {\color{black} G^{<}} - {\color{black} \mathcal{B}^{R}_{(1)}} {\color{black} G^{R}} {\color{black} \mathcal{B}^{R}_{(1)}} {\color{black} G^{<}} \mathcal{A}^{A}_{(0)} \\ \times {\color{black} G^{A}} + {\color{black} \mathcal{B}^{R}_{(1)}} {\color{black} G^{R}} {\color{black} \mathcal{B}^{R}_{(1)}} {\color{black} G^{R}} \partial_{\omega} \widetilde{\Sigma}^{<}_{{\color{black}  tot}(0)} {\color{black} G^{A}} + \mathcal{A}^{R}_{(0)} {\color{black} G^{R}} {\color{black} \mathcal{B}^{R}_{(1)}} {\color{black} G^{R}} {\color{black} \mathcal{B}^{R}_{(1)}} {\color{black} G^{<}} + \mathcal{A}^{R}_{(0)} {\color{black} G^{R}} {\color{black} \mathcal{B}^{R}_{(1)}} {\color{black} G^{<}} \\ \times {\color{black} \mathcal{B}^{A}_{(1)}} {\color{black} G^{A}} + \mathcal{A}^{R}_{(0)} {\color{black} G^{R}} {\color{black} \mathcal{B}^{R}_{(1)}} {\color{black} G^{R}} \partial_{T} \widetilde{\Sigma}^{<}_{{\color{black}  tot}(0)} {\color{black} G^{A}} \Big] \mathcal{A}^{A}_{(0)} {\color{black} G^{A}} - \frac{1}{4} {\color{black} G^{R}} \Big[ {\color{black} \mathcal{B}^{R}_{(1)}} {\color{black} G^{R}} \mathcal{A}^{R}_{(0)} {\color{black} G^{R}} \partial_{T \omega} \widetilde{\Sigma}^{R}_{{\color{black}  tot}(0)} \\ + \Big( \mathcal{A}^{R}_{(0)} {\color{black} G^{R}} \Big)^2 {\color{black} \mathcal{C}^{R}} \Big] {\color{black} G^{<}} - \frac{1}{4} {\color{black} G^{R}} \Big( {\color{black} \mathcal{B}^{R}_{(1)}} {\color{black} G^{R}} \mathcal{A}^{R}_{(0)} {\color{black} G^{<}} \partial_{T \omega} \widetilde{\Sigma}^{A}_{{\color{black}  tot}(0)} + \mathcal{A}^{R}_{(0)} {\color{black} G^{R}} \mathcal{A}^{R}_{(0)} {\color{black} G^{<}} \mathcal{C}^{A}_{\alpha \beta (2)} \Big) {\color{black} G^{A}} \\ - \frac{1}{4} {\color{black} G^{R}} \Big( {\color{black} \mathcal{B}^{R}_{(1)}} {\color{black} G^{R}} {\color{black} \mathcal{B}^{R}_{(1)}} {\color{black} G^{R}} \partial^2_{\omega} \widetilde{\Sigma}^{R}_{{\color{black}  tot}(0)} + \mathcal{A}^{R}_{(0)} {\color{black} G^{R}} {\color{black} \mathcal{B}^{R}_{(1)}} {\color{black} G^{R}} \partial_{T \omega} \widetilde{\Sigma}^{R}_{{\color{black}  tot}(0)} \Big) {\color{black} G^{<}} - \frac{1}{4} {\color{black} G^{R}} \Big( {\color{black} \mathcal{B}^{R}_{(1)}} {\color{black} G^{R}} {\color{black} \mathcal{B}^{R}_{(1)}} \\ {\color{black} G^{<}} \partial^2_{\omega} \widetilde{\Sigma}^{A}_{{\color{black}  tot}(0)} + \mathcal{A}^{R}_{(0)} {\color{black} G^{R}} {\color{black} \mathcal{B}^{R}_{(1)}} {\color{black} G^{<}} \partial_{T \omega} \widetilde{\Sigma}^{A}_{{\color{black}  tot}(0)} \Big) {\color{black} G^{A}} - \frac{1}{4} {\color{black} G^{R}} \Big( \mathcal{A}^{R}_{(0)} {\color{black} G^{R}} \mathcal{A}^{R}_{(0)} {\color{black} G^{<}} {\color{black} \mathcal{B}^{A}_{(1)}} {\color{black} G^{A}} {\color{black} \mathcal{B}^{A}_{(1)}} {\color{black} G^{A}} \\ - {\color{black} \mathcal{B}^{R}_{(1)}} {\color{black} G^{R}} \mathcal{A}^{R}_{(0)} {\color{black} G^{<}} {\color{black} \mathcal{B}^{A}_{(1)}} {\color{black} G^{A}} \mathcal{A}^{A}_{(0)} {\color{black} G^{A}} \Big) + \frac{1}{4} {\color{black} G^{R}} \Big[ \mathcal{A}^{R}_{(0)} {\color{black} G^{R}} {\color{black} \mathcal{B}^{R}_{(1)}} {\color{black} G^{<}} \mathcal{A}^{A}_{(0)} {\color{black} G^{A}} {\color{black} \mathcal{B}^{A}_{(1)}} {\color{black} G^{A}} - {\color{black} \mathcal{B}^{R}_{(1)}} {\color{black} G^{R}} \\ {\color{black} \mathcal{B}^{R}_{(1)}} {\color{black} G^{<}} \Big( \mathcal{A}^{A}_{(0)} {\color{black} G^{A}} \Big)^2 \Big] - \frac{1}{4} {\color{black} G^{R}} \Big[ {\color{black} \mathcal{B}^{R}_{(1)}} {\color{black} G^{R}} \mathcal{A}^{R}_{(0)} {\color{black} G^{R}} {\color{black} \mathcal{B}^{R}_{(1)}} \Big( - {\color{black} G^{R}} \mathcal{A}^{R}_{(0)} {\color{black} G^{<}} - {\color{black} G^{<}}  \mathcal{A}^{A}_{(0)} {\color{black} G^{A}} \\ + {\color{black} G^{R}} \partial_{\omega} \widetilde{\Sigma}^{<}_{{\color{black}  tot}(0)} {\color{black} G^{A}} \Big) + \mathcal{A}^{R}_{(0)} {\color{black} G^{R}} \mathcal{A}^{R}_{(0)} {\color{black} G^{R}} {\color{black} \mathcal{B}^{R}_{(1)}} \Big( {\color{black} G^{R}} {\color{black} \mathcal{B}^{R}_{(1)}} {\color{black} G^{<}} + {\color{black} G^{<}} {\color{black} \mathcal{B}^{A}_{(1)}} {\color{black} G^{A}} + {\color{black} G^{R}} \partial_{T} \widetilde{\Sigma}^{<}_{{\color{black}  tot}(0)} \\ {\color{black} G^{A}} \Big) \Big] + \frac{1}{4} {\color{black} G^{R}} \Big[ {\color{black} \mathcal{B}^{R}_{(1)}} {\color{black} G^{R}} {\color{black} \mathcal{B}^{R}_{(1)}} {\color{black} G^{R}} \mathcal{A}^{R}_{(0)} \Big( - {\color{black} G^{R}} \mathcal{A}^{R}_{(0)} {\color{black} G^{<}} - {\color{black} G^{<}} \mathcal{A}^{A}_{(0)} {\color{black} G^{A}} + {\color{black} G^{R}} \partial_{\omega} \widetilde{\Sigma}^{<}_{{\color{black}  tot}(0)} {\color{black} G^{A}} \Big) \\[0.5ex] + \mathcal{A}^{R}_{(0)} {\color{black} G^{R}} {\color{black} \mathcal{B}^{R}_{(1)}} {\color{black} G^{R}} \mathcal{A}^{R}_{(0)} \Big( {\color{black} G^{R}} {\color{black} \mathcal{B}^{R}_{(1)}} {\color{black} G^{<}} + {\color{black} G^{<}} {\color{black} \mathcal{B}^{A}_{(1)}} {\color{black} G^{A}} + {\color{black} G^{R}} \partial_{T} \widetilde{\Sigma}^{<}_{{\color{black}  tot}(0)} {\color{black} G^{A}} \Big) \Big],
\end{multline}
\begin{multline}
{\color{black} G^{R}} \hat{P}^{R}_{a(1)} A_3 = - \frac{1}{4} {\color{black} G^{R}} \Big[ \mathcal{A}^{R}_{(0)} {\color{black} G^{R}}, {\color{black} \mathcal{B}^{R}_{(1)}} {\color{black} G^{R}} \Big]_{-} \Big( {\color{black} \mathcal{B}_{(1)}^{R}} {\color{black} G^{R}} \partial_{\omega} \widetilde{\Sigma}^{<}_{{\color{black}  tot}(0)} {\color{black} G^{A}} + \mathcal{A}_{(0)}^{R} {\color{black} G^{R}} \partial_{T} \widetilde{\Sigma}^{<}_{{\color{black}  tot}(0)} {\color{black} G^{A}} \Big) \\ - \frac{1}{4} {\color{black} G^{R}} \Big( {\color{black} \mathcal{B}_{(1)}}^{R} {\color{black} G^{R}} \partial_{T \omega} \widetilde{\Sigma}^{R}_{{\color{black}  tot}(0)} {\color{black} G^{R}} \partial_{\omega} \widetilde{\Sigma}^{<}_{{\color{black}  tot}(0)} {\color{black} G^{A}} - {\color{black} \mathcal{B}_{(1)}^{R}} {\color{black} G^{R}} \partial^2_{\omega} \widetilde{\Sigma}^{R}_{{\color{black}  tot}(0)} {\color{black} G^{R}} \partial_{T} \widetilde{\Sigma}^{<}_{{\color{black}  tot}(0)} {\color{black} G^{A}} \Big) - \frac{1}{4} {\color{black} G^{R}} \\ \times \Big( \mathcal{A}^{R}_{(0)} {\color{black} G^{R}} {\color{black} \mathcal{C}^{R}} {\color{black} G^{R}} \partial_{\omega} \widetilde{\Sigma}^{<}_{{\color{black}  tot}(0)} {\color{black} G^{A}} - \mathcal{A}^{R}_{(0)} {\color{black} G^{R}} \partial_{T \omega} \widetilde{\Sigma}^{R}_{{\color{black}  tot}(0)} {\color{black} G^{R}} \partial_{T} \widetilde{\Sigma}^{<}_{{\color{black}  tot}(0)} {\color{black} G^{A}} \Big) + \frac{1}{4} {\color{black} G^{R}} \Big[ {\color{black} \mathcal{B}_{(1)}^{R}} {\color{black} G^{R}} \\ \times {\color{black} \mathcal{B}_{(1)}^{R}} {\color{black} G^{R}} \mathcal{A}_{(0)}^{R} {\color{black} G^{R}} \partial_{\omega} \widetilde{\Sigma}^{<}_{{\color{black}  tot}(0)} {\color{black} G^{A}} + {\color{black} \mathcal{B}_{(1)}^{R}} \Big( {\color{black} G^{R}} \mathcal{A}_{(0)}^{R} \Big)^2 {\color{black} G^{R}} \partial_{T} \widetilde{\Sigma}^{<}_{{\color{black}  tot}(0)} {\color{black} G^{A}} \Big] - \frac{1}{4} {\color{black} G^{R}} \Big[ \mathcal{A}^{R}_{(0)} {\color{black} G^{R}} {\color{black} \mathcal{B}_{(1)}^{R}} \\ \times {\color{black} G^{R}} {\color{black} \mathcal{B}_{(1)}^{R}} {\color{black} G^{R}} \partial_{\omega} \widetilde{\Sigma}^{<}_{{\color{black}  tot}(0)} {\color{black} G^{A}} + \Big( \mathcal{A}^{R}_{(0)} {\color{black} G^{R}} \Big)^2  {\color{black} \mathcal{B}_{(1)}^{R}} {\color{black} G^{R}} \partial_{T} \widetilde{\Sigma}^{<}_{{\color{black}  tot}(0)} {\color{black} G^{A}} \Big] - \frac{1}{4} {\color{black} G^{R}} \Big( {\color{black} \mathcal{B}_{(1)}^{R}} {\color{black} G^{R}} {\color{black} \mathcal{B}_{(1)}^{R}} \\ \times {\color{black} G^{R}} \partial^2_{\omega} \widetilde{\Sigma}^{<}_{{\color{black}  tot}(0)} {\color{black} G^{A}} + {\color{black} \mathcal{B}_{(1)}^{R}} {\color{black} G^{R}} \mathcal{A}_{(0)}^{R} {\color{black} G^{R}} \partial_{T \omega} \widetilde{\Sigma}^{<}_{{\color{black}  tot}(0)} {\color{black} G^{A}} \Big) - \frac{1}{4} {\color{black} G^{R}} \Big[ \mathcal{A}^{R}_{(0)} {\color{black} G^{R}} {\color{black} \mathcal{B}_{(1)}^{R}} {\color{black} G^{R}} \partial_{T \omega} \widetilde{\Sigma}^{<}_{T (0)} \\ \times {\color{black} G^{A}} + \Big( \mathcal{A}^{R}_{(0)} {\color{black} G^{R}} \Big)^2 \partial^2_{T} \widetilde{\Sigma}^{<}_{{\color{black}  tot}(0)} {\color{black} G^{A}} \Big] + \frac{1}{4} {\color{black} G^{R}} \Big( {\color{black} \mathcal{B}_{(1)}^{R}} {\color{black} G^{R}} {\color{black} \mathcal{B}_{(1)}^{R}} {\color{black} G^{R}} \partial_{\omega} \widetilde{\Sigma}^{<}_{{\color{black}  tot}(0)} {\color{black} G^{A}} \mathcal{A}_{(0)}^{A} {\color{black} G^{A}} \\ + {\color{black} \mathcal{B}_{(1)}^{R}} {\color{black} G^{R}} \mathcal{A}_{(0)}^{R} {\color{black} G^{R}} \partial_{T} \widetilde{\Sigma}^{<}_{{\color{black}  tot}(0)} {\color{black} G^{A}} \mathcal{A}_{(0)}^{A} {\color{black} G^{A}} \Big) - \frac{1}{4} {\color{black} G^{R}} \Big[ \mathcal{A}^{R}_{(0)} {\color{black} G^{R}} {\color{black} \mathcal{B}_{(1)}^{R}} {\color{black} G^{R}} \partial_{\omega} \widetilde{\Sigma}^{<}_{{\color{black}  tot}(0)} {\color{black} G^{A}} {\color{black} \mathcal{B}_{(1)}^{A}} {\color{black} G^{A}} \\ + \Big( \mathcal{A}^{R}_{(0)} {\color{black} G^{R}} \Big)^2 \partial_{T} \widetilde{\Sigma}^{<}_{{\color{black}  tot}(0)} {\color{black} G^{A}} {\color{black} \mathcal{B}_{(1)}^{A}} {\color{black} G^{A}} \Big]
\end{multline}
and 
\begin{multline}
{\color{black} G^{R}} \hat{P}^{R}_{a(1)} A_4 = \frac{1}{4} {\color{black} G^{R}} \Big[ \mathcal{A}^{R}_{(0)} {\color{black} G^{R}}, {\color{black} \mathcal{B}^{R}_{(1)}} {\color{black} G^{R}} \Big]_{-} \Big( \partial_{\omega} \widetilde{\Sigma}^{<}_{{\color{black} tot}(0)} {\color{black} G^{A}} {\color{black} \mathcal{B}^{A}_{(1)}} {\color{black} G^{A}} + \partial_{T} \widetilde{\Sigma}^{<}_{{\color{black} tot} (0)} {\color{black} G^{A}} \mathcal{A}^{A}_{(0)} {\color{black} G^{A}} \Big) \\ + \frac{1}{4} {\color{black} G^{R}} \Big( {\color{black} \mathcal{B}^{R}_{(1)}} {\color{black} G^{R}} \partial^2_{\omega} \widetilde{\Sigma}^{<}_{{\color{black} tot}(0)} {\color{black} G^{A}} {\color{black} \mathcal{B}^{A}_{(1)}} {\color{black} G^{A}} + {\color{black} \mathcal{B}^{R}_{(1)}} {\color{black} G^{R} } \partial_{T \omega} \widetilde{\Sigma}^{<}_{{\color{black} tot}(0)} {\color{black} G^{A}} \mathcal{A}^{A}_{(0)} {\color{black} G^{A}} \Big) + \frac{1}{4} {\color{black} G^{R}} \Big( \mathcal{A}^{R}_{(0)} \\ \times {\color{black} G^{R}} \partial_{T \omega} \widetilde{\Sigma}^{<}_{{\color{black} tot}(0)} G^{A} {\color{black} \mathcal{B}^{A}_{(1)}} {\color{black} G^{A}} + \mathcal{A}^{R}_{(0)} {\color{black} G^{R}} \partial^2_{T} \widetilde{\Sigma}^{<}_{{\color{black} tot}(0)} {\color{black} G^{A}} \mathcal{A}^{A}_{(0)} {\color{black} G^{A}} \Big) - \frac{1}{4} {\color{black} G^{R}} \Big[ {\color{black} \mathcal{B}^{R}_{(1)}} {\color{black} G^{R}} \partial_{\omega} \widetilde{\Sigma}^{<}_{{\color{black} tot}(0)} \\ \times {\color{black} G^{A}} \mathcal{A}^{A}_{(0)} {\color{black} G^{A}} {\color{black} \mathcal{B}^{A}_{(1)}} {\color{black} G^{A}} + {\color{black} \mathcal{B}^{R}_{(1)}} {\color{black} G^{R}} \partial_{T} \widetilde{\Sigma}^{<}_{{\color{black} tot}(0)} {\color{black} G^{A}} \Big( \mathcal{A}^{A}_{(0)} {\color{black} G^{A}} \Big)^2 \Big] + \frac{1}{4} {\color{black} G^{R}} \Big( \mathcal{A}^{R}_{(0)} {\color{black} G^{R}} \partial_{\omega} \widetilde{\Sigma}^{<}_{{\color{black} tot}(0)} {\color{black} G^{A}} \\ \times {\color{black} \mathcal{B}^{A}_{(1)}} {\color{black} G^{A}} {\color{black} \mathcal{B}^{A}_{(1)}} {\color{black} G^{A}} + \mathcal{A}^{R}_{(0)} {\color{black} G^{R}} \partial_{T} \widetilde{\Sigma}^{<}_{{\color{black} tot}(0)} {\color{black} G^{A}} {\color{black} \mathcal{B}^{A}_{(1)}} {\color{black} G^{A}} \mathcal{A}^{A}_{(0)} {\color{black} G^{A}} \Big) + \frac{1}{4} {\color{black} G^{R}} \Big( {\color{black} \mathcal{B}^{R}_{(1)}} {\color{black} G^{R}} \partial_{\omega} \widetilde{\Sigma}^{<}_{{\color{black} tot}(0)} \\ \times {\color{black} G^{A}} \partial_{T \omega} \widetilde{\Sigma}^{A}_{{\color{black} tot}(0)} G^{A} - {\color{black} \mathcal{B}^{R}_{(1)}} {\color{black} G^{R}} \partial_{T} \widetilde{\Sigma}^{<}_{{\color{black} tot}(0)} {\color{black} G^{A}} \partial^2_{\omega} \widetilde{\Sigma}^{A}_{{\color{black} tot} (0)} {\color{black} G^{A}} \Big) + \frac{1}{4} {\color{black} G^{R}} \Big( \mathcal{A}^{R}_{(0)} {\color{black} G^{R}} \partial_{\omega} \widetilde{\Sigma}^{<}_{{\color{black} tot}(0)} {\color{black} G^{A}} \\ \times {\color{black} \mathcal{C}^{A}} {\color{black} G^{A}} - \mathcal{A}^{R}_{(0)} {\color{black} G^{R}} \partial_{T} \widetilde{\Sigma}^{<}_{{\color{black}tot}(0)} \widetilde{G}^{A} \partial_{T \omega} \widetilde{\Sigma}^{A}_{{\color{black} tot} (0)} {\color{black} G^{A}} \Big) - \frac{1}{4} {\color{black} G^{R}} \Big[ {\color{black} \mathcal{B}^{R}_{(1)}} {\color{black} G^{R}} \partial_{\omega} \widetilde{\Sigma}^{<}_{{\color{black} tot}(0)} {\color{black} G^{A}} {\color{black} \mathcal{B}^{A}_{(1)}} {\color{black} G^{A}} \\ \times \mathcal{A}^{A}_{(0)} {\color{black} G^{A}} + {\color{black} \mathcal{B}^{R}_{(1)}} {\color{black} G^{R}} \partial_{T} \widetilde{\Sigma}^{<}_{{\color{black} tot}(0)} {\color{black} G^{A}} \Big( \mathcal{A}^{A}_{(0)} {\color{black} G^{A}} \Big)^2 \Big] + \frac{1}{4} {\color{black} G^{R}} \Big[ \mathcal{A}^{R}_{(0)} {\color{black} G^{R}} \partial_{\omega} \widetilde{\Sigma}^{<}_{{\color{black} tot}(0)} {\color{black} \Big( G^{A}} {\color{black} \mathcal{B}^{A}_{(1)} \Big)^2} \\ \times {\color{black} G^{A}} + \mathcal{A}^{R}_{(0)} {\color{black} G^{R}} \partial_{T} \widetilde{\Sigma}^{<}_{{\color{black} tot}(0)} {\color{black} G^{A}} \mathcal{A}^{A}_{(0)} {\color{black} G^{A}} {\color{black} \mathcal{B}^{A}_{(1)}} {\color{black} G^{A}} \Big].
\end{multline}
It follows that one can re-assemble ${\color{black} G^{R}} \hat{P}^{R}_{a(1)} \widetilde{G}_{(1)}^{<}$ through equation (\ref{combination}).

\newpage
\nocite{}
%\bibliography{/Users/dkosov/Documents/BIBLIOGRAPHY/thphys.bib}
%

\end{document}